\pdfoutput=1

\documentclass[10pt,conference]{IEEEtran}
\renewcommand{\baselinestretch}{0.99}
\IEEEoverridecommandlockouts
\usepackage{graphicx}
\usepackage{balance}  
\usepackage{enumitem}
\usepackage{hyperref}
\hypersetup{
    colorlinks,
    citecolor=red,
    filecolor=black,
    linkcolor=blue,
    urlcolor=red,
}
\usepackage{capt-of}
\usepackage{arydshln}
\usepackage{balance}

\begin{document}
\setlength\dashlinedash{0.1pt}
\setlength\dashlinegap{2pt}
\setlength\arrayrulewidth{0.5pt}


\title{Performance Benchmarking and Optimizing Hyperledger Fabric Blockchain Platform\vspace{-.3cm}}

\author{
%
%
	\IEEEauthorblockN{Parth Thakkar\textsuperscript{*}\thanks{\textsuperscript{*}Work done as part of undergraduate internship at IBM}}
	\IEEEauthorblockA{National Institute of Technology, Trichy, India\\
	 \emph{pthakker@in.ibm.com}} \and
	\IEEEauthorblockN{Senthil Nathan N}
        \IEEEauthorblockA{IBM Research Lab, India\\
	 \emph{snatara7@in.ibm.com}} \and
	\IEEEauthorblockN{Balaji Viswanathan}
	\IEEEauthorblockA{IBM Research Lab, India\\
	 \emph{bviswana@in.ibm.com}}
	 \vspace*{-1cm}
}

\vspace{-1cm}
\maketitle
\begin{abstract}
	The rise in popularity of permissioned blockchain platforms in recent time is significant. 
Hyperledger Fabric 
is one such permissioned blockchain platform and one of the Hyperledger projects hosted by the Linux Foundation \cite{linux-foundation}. The Fabric 
comprises of various components such as smart-contracts, endorsers, committers, validators, 
and orderers. As the performance of blockchain platform is a major concern for enterprise 
applications, in this work, we perform a comprehensive empirical study to characterize the 
performance of Hyperledger Fabric and identify potential performance bottlenecks to gain 
a better understanding of the system.
\par
We follow a two-phased approach. In the first phase, our goal is to understand the 
impact of various configuration parameters such as block size, endorsement policy, channels, 
resource allocation, state database choice on the transaction throughput \& latency to provide various
guidelines on configuring these parameters. In addition, we also aim to identify 
performance bottlenecks and hotspots. We observed that (1) endorsement policy verification, (2)
sequential policy validation of transactions in a block, and (3) state validation and commit (with CouchDB) 
were the three major bottlenecks. 
\par
In the second phase, we focus on optimizing Hyperledger Fabric v1.0 based 
on our observations.  We introduced and studied various simple 
optimizations such as aggressive caching for endorsement policy verification 
in the cryptography component (3$\times$ improvement in the performance) and parallelizing endorsement 
policy verification (7$\times$ improvement). Further, we enhanced and measured the effect of an
existing bulk read/write optimization for CouchDB during state validation \& commit phase
(2.5$\times$ improvement). 
By combining all three optimizations\footnote{These optimizations are successfully adopted in 
Hyperledger Fabric v1.1}, we improved the overall throughput by $16\times$
(i.e., from 140 tps to 2250 tps). 

\end{abstract}

\section{Introduction}
Blockchain technologies initially gained popularity as they were seen 
as a way to get rid of the intermediary and decentralize the system. 
Since then, blockchain has witnessed a growing interest from different domains 
and use cases. A blockchain is a shared, distributed ledger that records transactions 
and is maintained by multiple nodes in the network where nodes do not trust each other. 
Each node holds the identical copy of the ledger which is usually represented as a chain of blocks, 
with each block being a logical sequence of transactions. Each block 
encloses the hash of its immediate previous block, thereby guaranteeing the
immutability of ledger. 
\par
Blockchain is often hailed as a new breed of database systems, in essence
being a distributed transaction processing system where the nodes are not
trusted and the system needs to achieve Byzantine fault tolerance \cite{pbft}. 
Blockchain provides serializability, immutability, and cryptographic
verifiability without a single point of trust unlike a database system; 
properties that have triggered blockchain adoption in a wide variety of industries. 
\par
A blockchain network can be either permissionless or permissioned. In a
permissionless network or public network such as Bitcoin \cite{bitcoin}, Ethereum \cite{ethereum}, 
anyone can join the network to perform transactions. Due to a large number of nodes
in a public network, a proof-of-work consensus approach is used to order transactions and create
a block. In a permissioned network, the identity of each participant is known and
authenticated cryptographically such that blockchain can store who performed which 
transaction. In addition, such a network can have extensive access control mechanisms 
built-in to limit who can (a) read \& append to ledger data, (b) issue transactions, 
(c) administer participation in the blockchain network.
\par
A permissioned network is highly suitable for enterprise applications that 
require authenticated participants. Each node in a permissioned network can be owned by different 
organizations. Further, enterprise applications need complex data models and expressibility
which can be supported using \textit{smart-contracts} \cite{contracts}. 
Enterprises find value in being able to integrate diverse systems without having 
to build a centralized solution and to bring a level of trust among untrusting parties or to bring
in a trusted third-party. 
Trade Finance \cite{trade-finance} and Food Safety \cite{food-safety} are examples of blockchain
applications where participants see value in visibility advantages it offers as compared to the 
existing loosely coupled centralized systems.
\par
There is a lot of concern  about the performance of
permissioned blockchain platforms and their ability to handle a huge volume of transactions 
at low latency. Another concern is the richness of language to describe the transactions. 
Different blockchain platforms such as Quorum \cite{quorum}, Corda \cite{corda} address these concerns using different 
techniques derived from the distributed systems domain. Hyperledger Fabric \cite{fabcoin}
is an enterprise-grade open-source permissioned blockchain platform
which has a modular design and a high degree of specifiability through trust
models and pluggable components. Fabric is currently being used in many different
use cases such as Global Trade Digitization \cite{gtd}, SecureKey \cite{securekey}, Everledger \cite{everledger}
and is the focus of our performance study.
\par
Fabric consists of various components such as endorsers, ordering service, and committers.
Further, it constitutes various phases in processing a transaction such as endorsement phase, 
ordering phase, validation and commit phase. Due to numerous components and phases, Fabric 
provides various configurable parameters such as block size, endorsement policy, channels, 
state database. Hence, one of the main challenges in setting up an efficient blockchain network 
is finding the right set of values for these parameters. For e.g., depending on the
application and requirements, one might need to answer the following questions:
\begin{itemize}[itemsep=0.5pt, leftmargin=20pt]
	\item What should be the block size to achieve a lower latency?
	\item How many channels can be created and what should be the resource allocation?
	\item What types of endorsement policy is more efficient?
	\item How much is the performance difference between GoLevelDB \cite{golevel} and 
		CocuhDB \cite{couch} when it is used as the state database? 
\end{itemize}
To answer above questions and to identify the performance bottlenecks, we 
perform a comprehensive empirical study of Fabric v1.0 with various configurable 
parameters. Specifically, our three major contributions are listed below.
\begin{enumerate}[itemsep=0.5pt, leftmargin=20pt]
	\item We conducted a comprehensive empirical study of Fabric platform by varying values assigned to 
	the five major parameters by conducting over 1000s of experiments. As a result, we provide six
	guidelines on configuring these parameters to attain the maximum performance. 
	\item We identified three major performance bottlenecks: (i) crypto operations, (ii) serial validation
	of transactions in a block, and (iii) multiple REST API calls to CouchDB.
\item Introduced and studied three simple optimizations\footnote{Source code is available 
	in \url{https://github.com/thakkarparth007/fabric}} to improve the overall performance 
	by 16$\times$ (i.e., from 140 tps to 2250 tps) for a single channel environment.
\end{enumerate}
\par
The rest of the paper is organized as follows: \S \ref{background} presents the architecture of
Hyperledger Fabric. \S \ref{goal} briefly describes the
goals of our study while \S \ref{methodology} delve into the experimental setup
and workload characteristics. \S \ref{exp-results} and \S \ref{optimizations} present our core
contributions while \S \ref{related-work} describes related work. Finally, we conclude this paper
in \S \ref{conclusion}.

\section{Background: Hyperledger Fabric Architecture \& Configuration Parameters}
\label{background}
The Hyperledger Fabric is an implementation of
permissioned blockchain system which has many unique properties suited for
enterprise-class applications. It can run arbitrary smart contracts
(\textit{a.k.a} chaincodes \cite{chaincode}) implemented in Go/JAVA/Nodejs language. It supports an application specifiable
trust model for transaction validation and a pluggable consensus protocol to
name a few. A Fabric network consists of different types of entities, peer nodes,
ordering service nodes and clients, belonging to different organizations. Each
of these has an identity on the network which is provided by a Membership
Service Provider (MSP) \cite{msp}, typically associated with an organization. All
entities in the network have visibility to identities of all organizations and
can verify them.
\subsection{Key Components in Fabric}
\begin{figure}
  \begin{center}
    \includegraphics[scale=0.33]{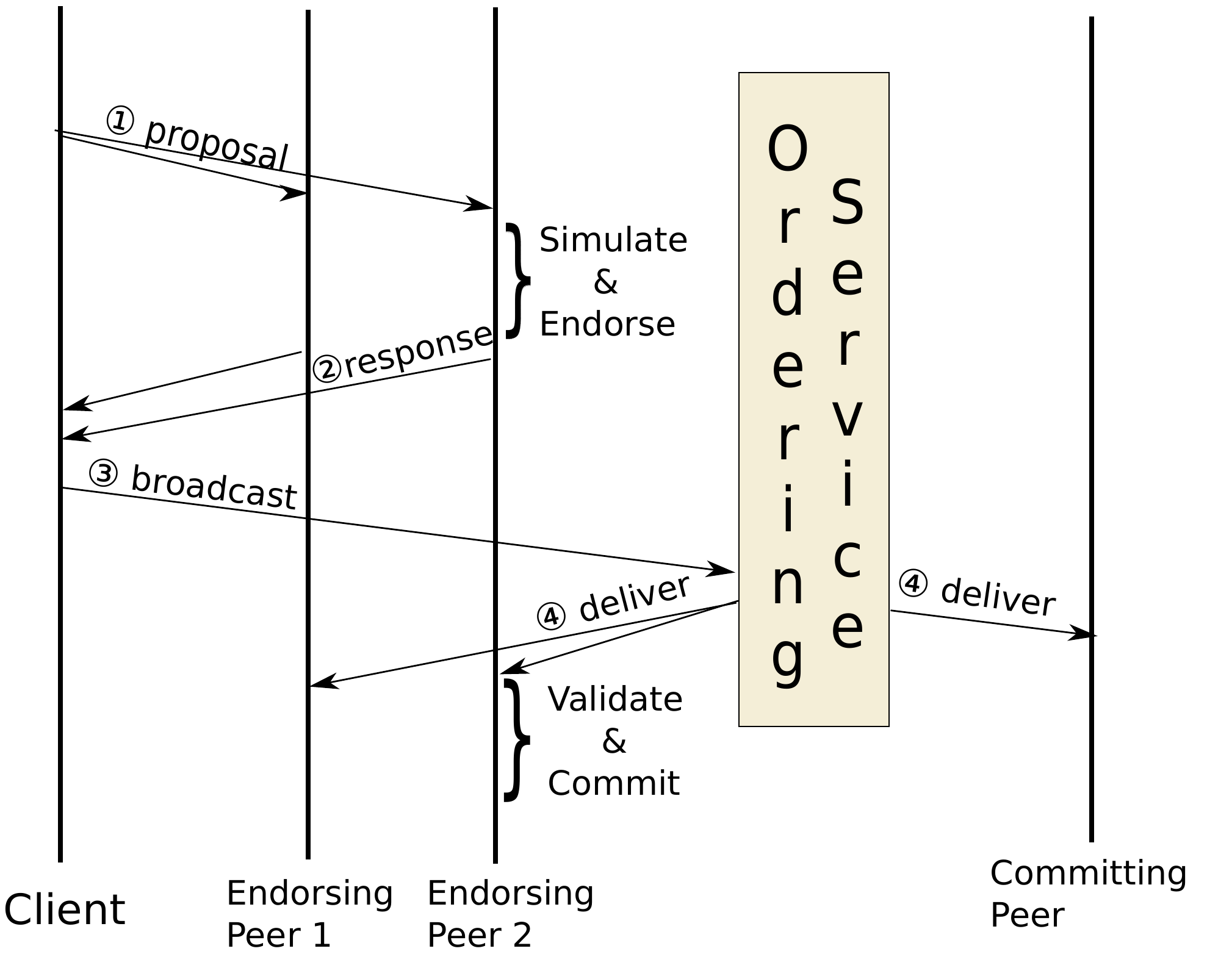}
    \vspace{-.3cm}
    \caption{Transaction flow.}
  \label{fig:flow}
  \end{center}
  \vspace{-.7cm}
\end{figure}
\par
\textbf{Peer.}
A peer node executes the chaincode, which implements a user smart-contract, and maintains 
the ledger in a file system. The chaincode is allowed access to the shared state by well-defined ledger APIs. A peer is
further segregated as an endorsing peer, one which has the chaincode logic
and executes it to endorse a transaction or a committing peer, one which does not 
hold the chaincode logic. Irrespective of this differentiation, both types of peer maintain the
ledger. Additionally, both peers maintain the current state as \textit{StateDB} in
a key-value store such that chaincode can query or modify the state using the database
query language. 
\par
\textit{Endorsement Policies.}
Chaincodes are written in general-purpose languages that execute on untrusted
peers in the network. This poses multiple problems, one of non-deterministic
execution and the other of trusting the results from any given peer. The
endorsement policy addresses these two concerns, by specifying as part of an
endorsement policy, the set of peers that need to simulate the transaction and 
endorse or digitally sign the execution results. 
Endorsement policies\footnote{\texttt{AND}('Org1.member', 'Org2.member', 'Org3.member') 
requests a signature from each of the three organization while \texttt{OR}('Org1.member',
\texttt{AND} ( 'Org2.member', 'Org3.member' )) requests either a signature from organization 1 
or two signatures, i.e., from both organization 2 \& 3.} are specified 
as boolean expressions over network principals identities. A principal here is a member of a specific
organization. 
\begin{figure}
  \begin{center}
    \includegraphics[scale=0.44]{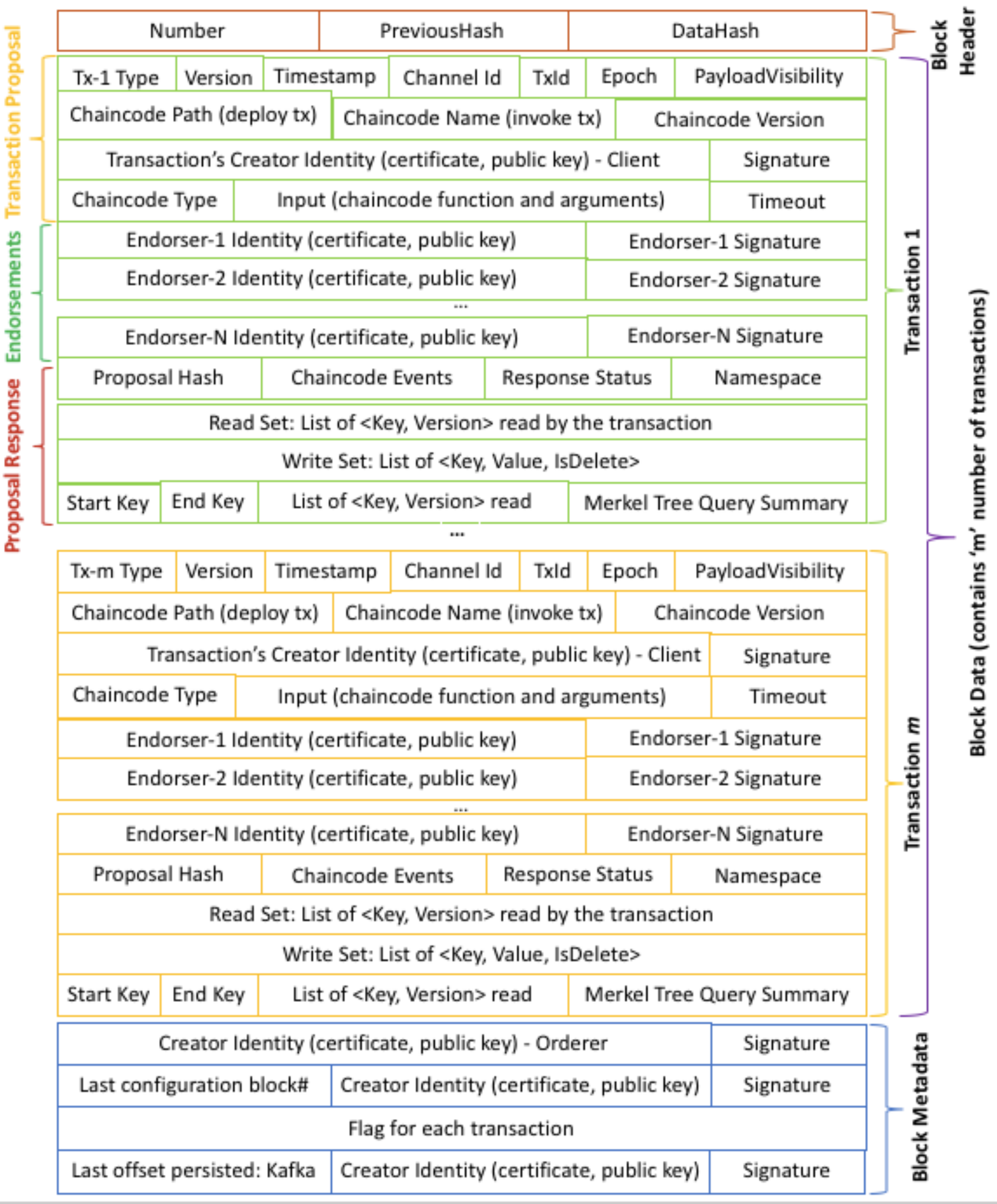}
    \vspace{-.3cm}
    \caption{Block Structure in Hyperledger Fabric v1.0}
  \label{fig:block-structure}
  \end{center}
  \vspace{-.7cm}
\end{figure}
\par
\textit{System chaincodes.}
System chaincode has the same programming model as normal user chaincodes and is built into the
peer executable, unlike user chaincodes. Fabric implements various system chaincodes; the life cycle
system chaincode (LSCC)---to install/instantiate/update chaincodes; the endorsement system chaincode (ESCC)---to 
endorse a transaction by digitally signing the response; the validation system chaincode (VSCC)---to validate 
a transaction's endorsement signature set against the endorsement policy; the configuration
system chaincode (CSCC) -- to manage channel configurations.
\par
\textit{Channel.}
Fabric introduces a concept called channel as a ``private'' subnet of communication between
two or more peers to provide a level of isolation. Transactions on a
channel are only seen by the peer members and participants. The immutable ledger 
and chaincodes are on a per-channel basis. Further, the consensus is applicable on a
per-channel basis, i.e., there is no defined order for
transaction across channels.
\par
\textbf{Ordering Service.}
An Ordering Service Node (OSN), participate in the consensus protocol and
\textit{cuts} block of transactions which is delivered to the peers by a gossip 
communication protocol. The structure of a block in Fabric v1.0 is shown in Figure \ref{fig:block-structure}. 
The ordering service is modular and
supports a pluggable consensus mechanism. By default, a serial ordering (i.e., consensus) 
is achieved using an underlying Kafka/Zookeeper cluster.
OSNs publish transactions to kafka topics and leverages the ordered and
immutable nature of records in kafka topic to generate a unique ordered
sequence of transactions in a block.  
A block is 
cut, when either a maximum number of new transactions, since the last block cut, are added added 
or a configured timeout since the last block cut has occurred.
When any one condition is satisfied, an OSN, publishes a
time-to-cut marker and cut a block of all transactions message offsets since
the last time-to-cut message offset. The block is then delivered to the peer
nodes.
\par
\textbf{Client.}
The client application is responsible for putting together a transaction
proposal as shown in Figure \ref{fig:block-structure}. The client submits the transaction proposal 
to 1-or-more peers simultaneously for collecting proposal responses with endorsements 
to satisfy the endorsement policy. It then \textit{broadcasts} the transaction
to the orderer to be included into a block and delivered to all peers for validation
and commit. In Fabric v1.0, the onus is on the client to
ensure that the transaction is well-formed and satisfies the endorsement
policies.
\subsection{Transaction Flow in Hyperledger Fabric}
\label{transaction-flow}
Unlike other Blockchain network which employ an order-execute \cite{order-execute} 
transaction model, the Fabric employs a simulate-order-validate \& commit model. 
Figure \ref{fig:flow} depicts the transaction flow which involves 3 steps, 
1) Endorsement Phase -- simulating the transaction on
selective peers and collecting the state changes; 2) Ordering Phase -- ordering
the transactions through a consensus protocol; and 3) Validation Phase --
validation followed by commit to ledger.
Before transactions can be submitted on Fabric, the network needs to be
bootstrapped with participating organizations, their MSPs and identities for
peers. First, a channel is created on the orderer network with respective 
organization MSPs. Second, peers of each organization join the channel and
initializes the ledger. Finally,
the required chaincodes are installed on the channel. 
\par
\textbf{Endorsement Phase.}
A client application using the Fabric SDK \cite{sdk-java, sdk-go, sdk-node}, constructs a transaction proposal
to invoke a chaincode function which in-turn will perform read and/or write
operations on the ledger state. 
The proposal is signed with the client's credentials and the
client sends it to 1-or-more endorsing peers simultaneously. The endorsement policy for the
chaincode dictates the organization peers the client needs to send the
proposal to for simulation.
\par
First, each endorsing peer verifies that the submitter is authorized  
to invoke transactions on the channel. Second, the peer 
executes the chaincode, which can access the current ledger state on peer. The transaction 
results include response value, read-set and write-set. All reads read the current state 
of ledger, but all writes are intercepted and modify a private transaction workspace. 
Third, the endorsing peer calls a system chaincode called ESCC which signs this 
transaction response with peer's identity and replies back to client with proposal response. 
Finally, the client inspects the proposal response to verify that it bears the signature of the peer. 
The client collects enough proposal response from different peers, verifies that the
endorsements are same. Since each peer could have executed the transaction
at different \textit{height} in the blockchain, it is possible that the proposal
response differs. In such cases, the client has to re-submit the proposal to
other peers, to obtain sufficient matching responses. 
\par
\textbf{Ordering Phase.}
The client \textit{broadcasts} a well-formed transaction message to the
Ordering Service. The transaction will contain the read-write sets, the
endorsing peer signatures and the Channel ID. The ordering service does not
need to inspect the contents of the transaction to perform its operation. It
receives transactions from different clients for various channels and
\textit{enqueues} them on a per-channel basis. It creates blocks of transactions
per channel, sign the block with its identity and \textit{delivers} them to 
peers using gossip messaging protocol.
\par
\textbf{Validation Phase.}
All peers, both endorsing and committing peers on a channel receive blocks from
the network. The peer first verifies the Orderer's signature on the block. 
Each valid block is decoded
and all transactions in a block goes through VSCC validation first before
performing MVCC validation.
\par
\textit{VSCC Validation.}
A Validation system chaincode evaluates endorsements in the transaction against 
the endorsement policy specified for the chaincode. If the endorsement policy is 
not satisfied, then that transaction is marked invalid.
\par
\textit{MVCC Validation.}
As the name implies, the Multi-Version Concurrency Control \cite{mvcc} check ensures that
the versions of keys read by a transaction during the endorsement phase are same as their current state in
the local ledger at commit time. This is similar to a read-write conflict check done for
concurrency control, and is performed sequentially on all the valid
transactions in the block (as marked by VSCC validation). If the read-set versions 
do not match, denoting that a concurrent previous (as-in earlier in this block or before) 
transaction modified the data read and was since (it's endorsement) successfully 
committed, the transaction is marked invalid. To ensure that no phantom reads occur, 
for range queries, the query is re-executed and compares the hashes of results 
(which is also stored as part of read-set captured during endorsement).
\par
\textbf{Ledger Update Phase.}
As the last step of transaction processing, the ledger is updated by appending
the block to the local ledger. The \textit{StateDB}, which holds the current
state of all keys is updated with the write-sets of valid transactions (as marked by MVCC validation). These
updates to the StateDB are performed atomically for a block of transactions and
applies the updates to bring the StateDB to the state after all transaction in
the block have been processed. 

\subsection{Configuration Parameters}
\label{config-parameters}
Our goal is to study the performance of Fabric under various conditions to
understand how choices of different facets of the system affect performance.
However, the parameter space is wide and we limit our choices to
comprehensively cover a few components and look widely at other aspects of the
system so that we can identify interplay of component level choices. To this
end, we choose to understand and characterize the overall performance primarily
from a peer's perspective. More specifically, we keep the Orderer, Gossip
(physical network) etc. static so that it does not affect our experiments and
observations. Next, we describe the five parameters considered in this study 
and their significance. 
\par
\textbf{1) Block Size.} Transactions are batched at the orderer and
delivered as a block to peers using a gossip protocol. Each peer
processes one block at a time. Cryptographic processing like
orderer signature verification is done per-block unlike transaction
endorsement signatures verification, which is per-transaction.  Varying
blocksize also brings in the throughput-vs-latency tradeoff and for a better
picture, we study it in conjunction with the transaction arrival rate.
\par
\textbf{2) Endorsement Policy.} An endorsement policy dictates how many
executions of a transaction and signing need to happen before a transaction
request can be submitted to the orderer so that the transaction can pass the
VSCC validation phase at peers. The VSCC validation of a transaction' endorsements require
evaluation of endorsement policy expression against the collected
endorsements and checking for satisfiability \cite{sat}, which is NP-Complete.
Additionally, a check includes verifying that the identity and its signature.
The complexity of the endorsement policy will affect resources and the time taken
to collect and evaluate it.  
\par
\textbf{3) Channel.} Channels isolate transactions from
one another. Transactions submitted to different channels are ordered, delivered and
processed independent of each other, albeit on same peers. Channels bring
inherent parallelism to various aspects of transaction processing in the
Fabric. While number of channels to use, and what channels to transact on is
determined by the application and participant combinatorics, it has significant 
implications on platform performance and scalability.  
\par
\textbf{4) Resource Allocation.} Peers run CPU-intensive signature
computation and verification routines as part of system chaincodes. User
chaincodes executed by endorsing peers during transaction simulation add to
this mix. We vary the number of CPU cores on peer nodes to study
its effect. While network characteristics are important, we assume a datacenter
or high bandwidth network with very low latency for this study.
\par
\textbf{5) Ledger Database.} Fabric supports two alternatives for
key-value store, CouchDB \cite{couch} and GoLevelDB \cite{golevel} to maintain the current
state. Both are key-value stores, while GoLevelDB is an embedded database,
CouchDB uses a client-server model (accessed using REST API over a secure HTTP) and supports document/JSON data-model.  

\section{Problem Statement}
\label{goal}
The two primary goals of our work are:
\par
		\textbf{1) Performance Benchmarking.} To conduct an in-depth study of Fabric core components and benchmark Fabric 
	performance for common usage patterns. We aim to study the throughput and latency 
	characteristics of the system when varying the configuration of parameters 
		listed \S \ref{config-parameters} to understand the relationship between the performance metrics and parameters. 
	Based on our observations, we aim to derive and present a few high-level guidelines,
	which would be valuable to developers and deployment engineers. 
\par
		\textbf{2) Optimization.} To identify bottlenecks using code-level instrumentation and to 
	draw out action items to improve the overall performance of Fabric. On 
	identifying bottlenecks, our goal is to introduce and implement optimizations 
	to alleviate these bottlenecks.  
\begin{figure}
    \begin{center}
    \includegraphics[scale=0.22]{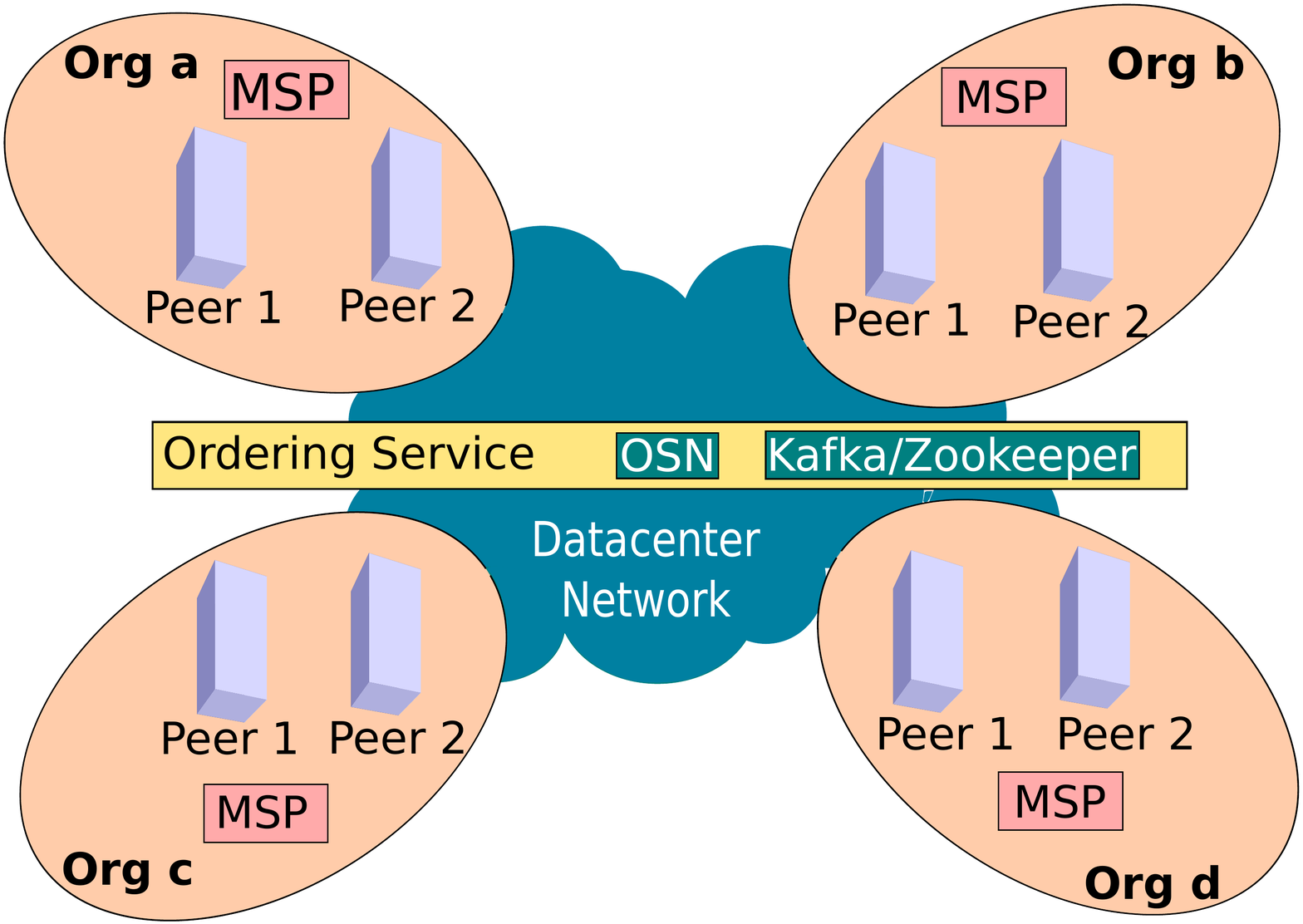}
	\vspace{-0.4cm}
    \caption{Experimental Setup}
   \label{fig:setup-diagram}
 \end{center}
	\vspace{-0.7cm}
\end{figure}

\section{Experimental Methodology}
\label{methodology}
We study the throughput and latency as the primary performance metrics for
Fabric.  \textbf{\textit{Throughput}} is the rate at which transactions are 
committed to ledger. \textbf{\textit{Latency}} is the time taken from
application sending the transaction proposal to the transaction commit and is made
up of the following latencies:
\begin{itemize}[nolistsep, leftmargin=10pt]
	\item \textbf{\textit{Endorsement latency}} -- the time taken for the client to collect all
proposal responses along with the endorsements.
	\item \textbf{\textit{Broadcast latency}} -- the time delay between client submitting to orderer 
	and orderer acknowledges the client. 
\item \textbf{\textit{Commit latency}} -- the time taken for the peer to validate and commit the
transaction. 
		\item \textbf{\textit{Ordering latency}} -- the time transaction spent on the ordering service. 
		As the performance of ordering service is not studied in this work, we are not 
		presenting this latency. 
\end{itemize}
Further, we define the following three latency at block level:
\begin{itemize}[nolistsep, leftmargin=10pt]
	\item \textbf{\textit{VSCC validation latency}} -- the time taken to validate all 
		transactions' endorsement signature set (in a block) against the endorsement policy.
	\item \textbf{\textit{MVCC validation latency}} -- the time taken to validate all
		transactions in a block by employing multi-version concurrency control as described in
		\S \ref{transaction-flow}.
	\item \textbf{\textit{Ledger update latency}} -- the time taken to update the
		state database with write-set of all valid transactions in a block.
\end{itemize}
Since one of the major goal of this work is to identify performance bottlenecks, our load
generator\footnote{\url{https://github.com/thakkarparth007/fabric-load-gen}} spans multiple clients 
each stresses the system by continuously generating transactions instead of following a
distribution model (say, Poisson). Each client also sends proposal requests in parallel 
and collates endorsements.  The transactions are submitted asynchronously to
achieve the specified rate without waiting for commits. However, the benchmark
framework tracks commit using our tools named \textit{fetch-block}\footnote{\url{https://github.com/cendhu/fetch-block}} 
to calculate throughput and latency. 
Further, we instrumented the Fabric source code to collect fine-grained latency such as MVCC latency 
and others latencies. For multi-channel experiments, all organizations and all its 
peers join the channel. While other combinations are possible, we believe our approach will stress test
the system. 

\subsection{Setup and Workloads}
\label{setup}
\begin{table}
 \caption{Default configuration for all experiments unless specified otherwise.}
	\vspace{-.3cm}
  \label{table:config-default}
  \begin{tabular}{  p{3.2cm} | p{4.4cm} }

        \hline
	  \textbf{Parameters} & \textbf{Values} \\ \hline
	  Number of Channels & 1 \\ \hdashline
	  Transaction Complexity & 1 KV write (1-w) of size 20 bytes \\ \hdashline
	  StateDB Database & GoLevelDB \\ \hdashline
	  Peer Resources & 32 vCPUs, 3 Gbps link \\ \hdashline
	  Endorsement Policy \texttt{AND}/\texttt{OR} & \texttt{OR} [\texttt{AND}(a ,b, c), \texttt{AND}(a, b, d), \texttt{AND}(b, c, d), \texttt{AND}(a, c, d)] \\ \hdashline
	  Block Size & 30 transactions per block \\ \hdashline
	  Block Timeout & 1 second\\ \hline
  \end{tabular}
	\vspace{-.4cm}
\end{table}

Our test Fabric network consists of 4 organizations, each with 2 endorsing peers
for a total of 8 peer nodes as depicted in Figure \ref{fig:setup-diagram}. There is 1 orderer node with a kafka-zookeeper
cluster backing it. All nodes and kafka-zookeeper run on the x86\_64 virtual machines in a IBM SoftLayer Datacenter.
Each virtual machine is allocated 32 vCPUs of Intel(R) Xeon(R) CPU E5-2683 v3 @ 2.00GHz
and 32 GB of memory. The three powerful client machines used to generate load was allocated with 56 vCPUs and 128 GB 
memory. Nodes are connected to the 3 Gbps Datacenter network. 

\par
In the lack of standard benchmarks for Blockchain, we built our own benchmarks
by surveying around 12 internal customer solutions built on Fabric for diverse
use cases. We identified common defining patterns, data models, and requirements
across the board. One of the recurrent pattern is for each chaincode
invocation to operate on exactly one asset or unit of data with the identifier
being passed to it. The query logic is done by higher level application layers,
often without querying blockchain data.  This pattern is modelled as simple
write-only transactions (1w, 3w and 5w - denoting number of keys written) in
our benchmark. Another common pattern is for a chaincode to read-and-write a
small set of keys, like read a JSON document, update a field and write it back.
We model these as read-writes of 1, 3 and 5 keys. As we have modeled our benchmark
to imitate real world blockchain applications in production, we have not 
considered other macro benchmarks.

\section{Experimental Results}
\label{exp-results}
In this section, we study the impact of various configurable parameters listed in \S \ref{config-parameters} 
on the performance of Hyperledger Fabric. 
The throughput and transaction latency presented in this section are averaged over multiple
runs. In total, we conducted more than 1000s of experiments. 
\subsection{Impact of Transaction Arrival Rate and Block Size}
\begin{table}
 \caption{Configuration to identify the impact of block size and transaction arrival rate.}
	\vspace{-.3cm}
  \label{table:config-blocksize}
  \begin{tabular}{  p{3cm} | p{4.8cm} }
        \hline
	  \textbf{Parameters} & \textbf{Values} \\ \hline
	  Tx. Arrival Rate & 25, 50, 75, 100, 125, 150, 175 (tps)\\ \hdashline
	  Block Size & 10, 30, 50, 100 (\#tx)\\ \hline
  \end{tabular}
\end{table}

\begin{figure}
    \begin{center}
     \includegraphics[scale=0.39]{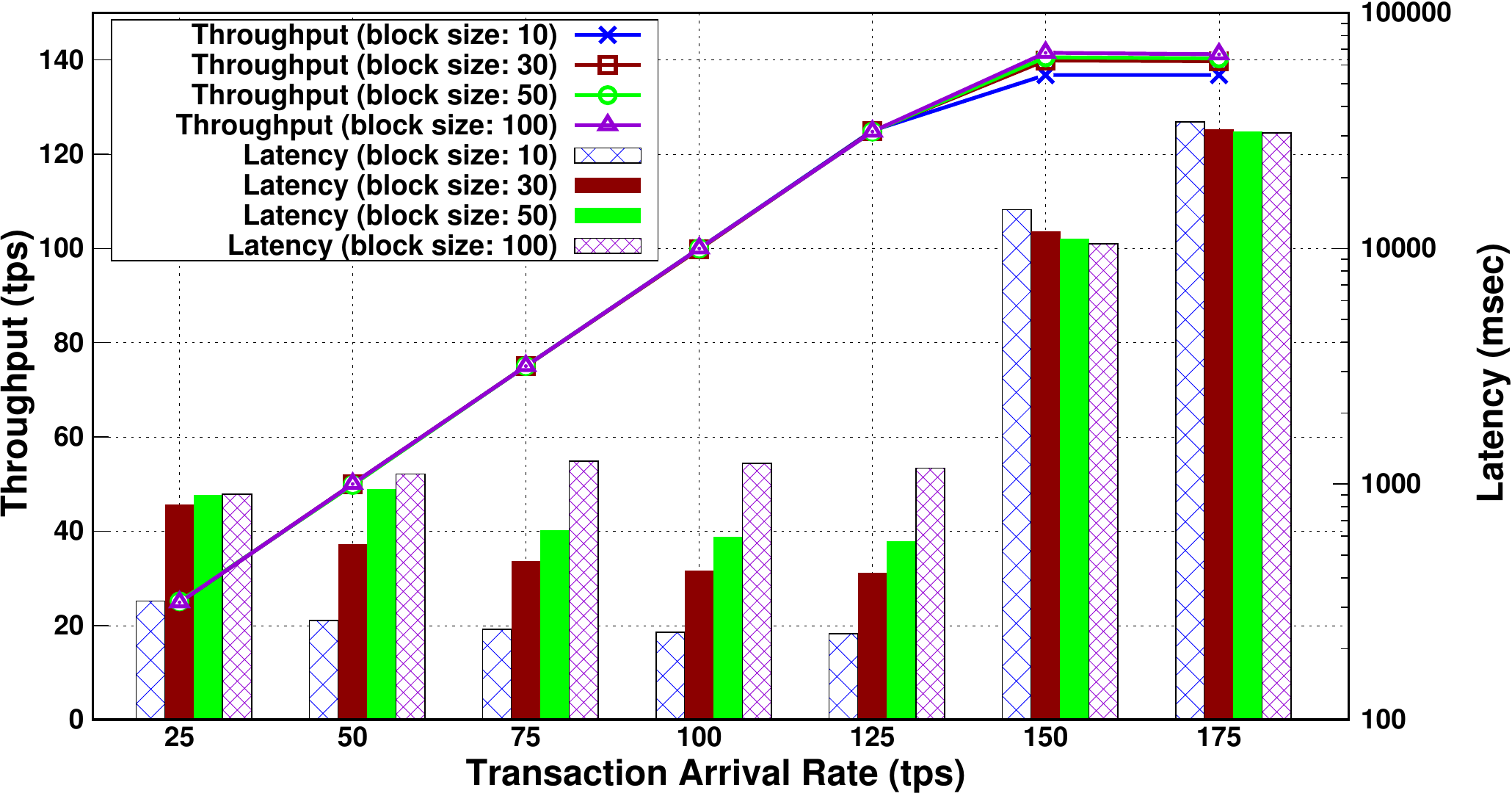}
     \vspace{-.3cm}
     \caption{Impact of the block size and transaction arrival rate on performance.}
   \label{fig:block-size}
 \end{center}
  \vspace{-.4cm}
\end{figure}

\par
Figure \ref{fig:block-size} plots the average throughput and latency for various block sizes
over different transaction arrival rates. Table \ref{table:config-blocksize} presents various 
transaction arrival rates and block sizes used. For other parameters, refer to 
Table \ref{table:config-default}.
\par
\begin{figure}[h]
    \begin{center}
     \includegraphics[scale=0.4]{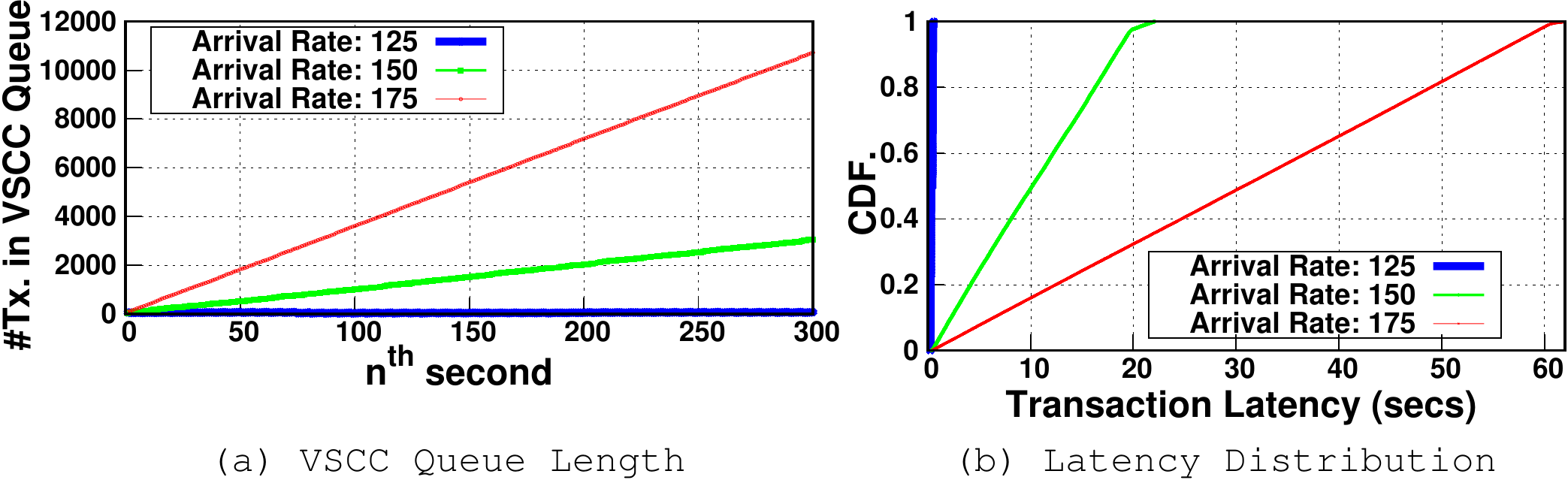}
     \vspace{-.3cm}
	    \caption{The length of VSCC queue, and latency distribution for various arrival rates
	    (block size was set to 30).}
   \label{fig:queue-length-block-size}
 \end{center}
  \vspace{-.3cm}
\end{figure}

\textbf{Observation 1:} \textit{With an increase in transaction arrival rate, the throughput 
increased linearly as expected till it flattened out at around 140 tps, the saturation point} as shown in Figure \ref{fig:block-size}. 
\textit{When the arrival rate was close to or above the saturation point, the latency increased 
significantly (i.e., from an order of 100s of ms to 10s of seconds)}. This is because
the number of ordered transactions waiting in the VSCC queue during validation phase grew rapidly 
(refer to Figure \ref{fig:queue-length-block-size}) which affected the commit latency. However, with 
further increase in the arrival rate, we observed no impact on the endorsement and broadcast 
latency but commit latency. This is because VSCC utilized only a single vCPU and hence 
new transaction proposals utilized other vCPUs on the peer for simulation and endorsement. As a result, only 
the validation phase became a bottleneck. In this experiment, the endorsement and broadcast latency was 
around 12 ms and 9 ms, respectively.

\par
\textbf{Observation 2:} \textit{For an arrival rate lower than the saturation point, with an increase 
in the block size, the latency increased}.
For e.g., when the
arrival rate was 50 tps, with an increase in the block size from 10 to 100, the transaction latency
increased 5-fold, from 242 ms to 1250 ms. The reason is that with an increase in the block size, the block creation 
time at the orderer increased and hence, on average, a transaction had to wait at the orderer for a little 
longer. For e.g., when the transaction arrival rate was 100 tps, for the block size of 50 and 100, 
the block creation rate was 2 and 1 block(s) per second, respectively, 
causing latency to double.
\par
\textbf{Observation 3:} \textit{For an arrival rate greater than the saturation point, with an increase 
in the block size, the latency decreased.} 
For e.g., when the arrival rate was 150 tps, with an increase in block size from 10 to 200, 
the transaction latency decreased from 14 secs to 10 secs. This is because the time taken to validate 
and commit a block of size $n$ was always lesser than the time taken to validate and commit  
$m$ blocks each of size $\frac{n}{m}$. 
As a result, throughput also increased by 3.5\%. Note that the block creation rate
at orderer node was always greater than the processing rate at validator irrespective of block size
and arrival rate.
\par
\textbf{Observation 4:} \textit{For a block size, the latency increases as arrival rate increases below 
the block size as the threshold\footnotetext{arrival rate adjusted to block timeout}. The latency decreases as arrival 
rate increases above the block size}.
For lower block sizes and at higher arrival rates, blocks were created faster (rather than waiting for a block timeout) which reduced 
the transaction waiting time at the orderer node.
In contrast, for instance, when the block size was 100 and arrival rate increased from 25 to 75 tps, the
latency increased from 900 ms to 1250 ms. The reason is that with the increase
in rate, the number of transactions in a block increased and so did the time
taken by validation and commit phase. 
Note that if block size limit was not reached within a second, a block was created
due to a block timeout.
\par
\textbf{Observation 5:} \textit{Even at the peak throughput, the resources utilization was very low}.
With an increase in the arrival rate from 25 to 175 tps, the avg CPU utilization merely increased from 1.4\% to 6.7\%\footnote{Unless specified otherwise,
CPU utilization are specified as an average across 32 vCPUs. In absolute terms, 6.7\% is equal to $6.7 \times 32 = 214\%$}. 
The reason is that the CPU intensive task performed during VSCC validation phase of a block (i.e., 
verification of signatures set against endorsement policy for each transaction) processed only one transaction at 
a time. Due to this serial execution, only one vCPU was utilized. 
\par
\textbf{Guideline 1}: When the transaction arrival rate is expected to be lower than the saturation point,
to achieve a lower transaction latency for applications, always use a lower block size. 
In such cases, the throughput will match the arrival rate.
 \par
\textbf{Guideline 2}: When the transaction arrival rate is expected to be high,
to achieve a higher throughput and a lower transaction latency, always use a higher block size.  
\par
\textbf{Action Item 1}: CPU resources are under-utilized. A potential optimization would be to process 
multiple transactions at a time during the VSCC validation phase as shown in \S \ref{pvscc}.
\begin{table}
 \caption{Configuration to identify the impact of endorsement policies.}
	\vspace{-.3cm}
  \label{table:config-endorsement}
  \begin{tabular}{  p{2.2cm} | p{5.6cm} }
        \hline
	  \textbf{Parameters} & \textbf{Values} \\ \hline
	  {Endorsement Policy} (\texttt{AND}/\texttt{OR}) &
	  		\vspace{-.2cm}
			\begin{enumerate}[nolistsep, leftmargin=10pt]
				\item \texttt{OR} [a, b, c, d]
				\item \texttt{OR} [\texttt{AND}(a ,b), \texttt{AND}(a, c), \texttt{AND}(a, d), \texttt{AND}(b, c), \texttt{AND}(b, d), \texttt{AND}(C, D)] 
				\item \texttt{OR} [\texttt{AND}(a ,b, c), \texttt{AND}(a, b, d), \texttt{AND}(b, c, d), \texttt{AND}(a, c, d)] 
				\item \texttt{AND} [a, b, c, d] 
			\end{enumerate} 
			\\ \hdashline
			  {Endorsement Policy} (\texttt{NOutOf}) & 
	  		\vspace{-.2cm}
			  \begin{enumerate}[nolistsep, leftmargin=10pt]
				  \item \texttt{1-OutOf} [a ,b ,c, d]  
				  \item \texttt{2-OutOf} [a ,b ,c, d]
				  \item \texttt{3-OutOf} [a ,b ,c, d]
				  \item \texttt{4-OutOf} [a ,b ,c, d] 
			  \end{enumerate} 	  		
			\\ \hdashline

	  {Tx. Arrival Rate} & 125, 150, 175 (tps) \\ \hline
  \end{tabular}
\end{table}

\subsection{Impact of Endorsement Policy}
\begin{figure}
    \begin{center}
     \includegraphics[scale=0.4]{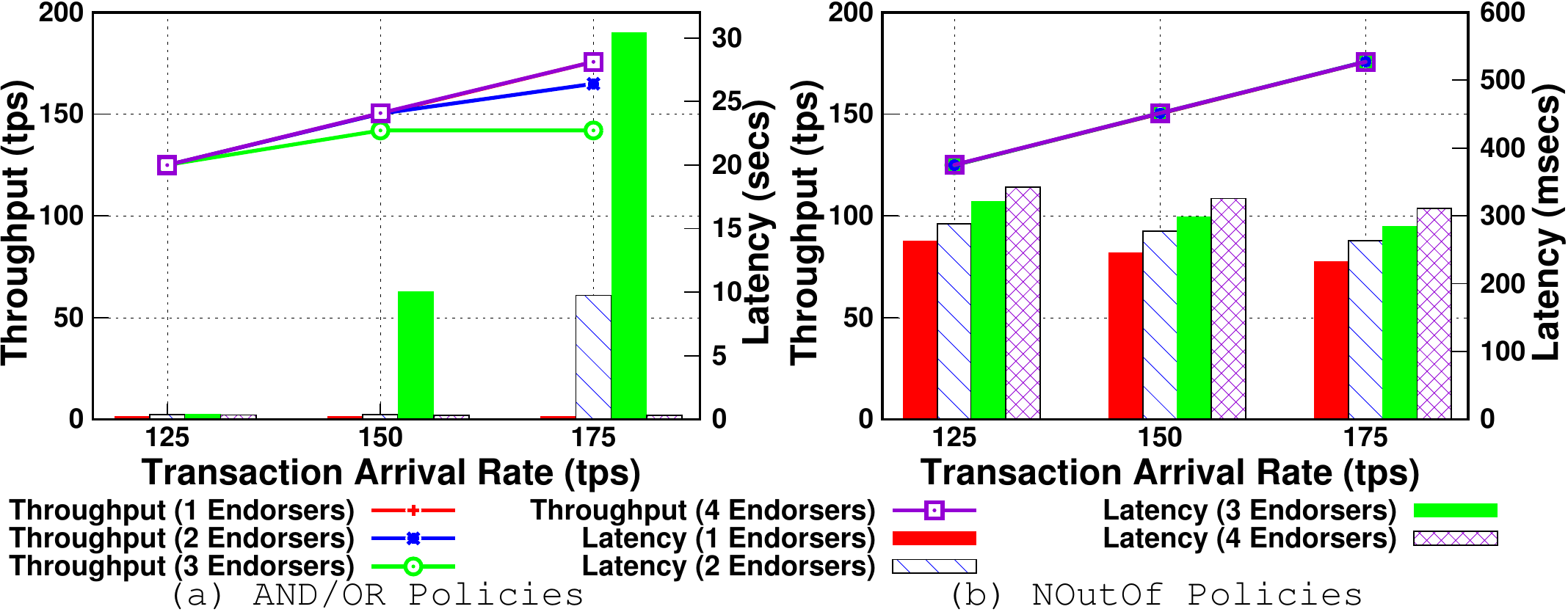}
     \vspace{-.3cm}
	    \caption{Impact of different endorsement policies (\texttt{AND/OR} and
	    \texttt{NOutOf}).}
   \label{fig:endorsement}
 \end{center}
  \vspace{-.6cm}
\end{figure}

%
Figure \ref{fig:endorsement} plots the throughput and latency for various endorsement policies defined  
using both \texttt{AND/OR} and \texttt{NOutOf} syntax over different transaction arrival rates. 
Table \ref{table:config-endorsement} presents various policies used in this study. Note that `a', `b', `c' and `d'
denotes four different organizations. For other parameters, refer to Table \ref{table:config-default}. Though the 
syntax (\texttt{AND/OR}, \texttt{NOutOf})
used to define the four endorsement policies are different, semantically they are same. 
For e.g., 3$^{rd}$ policy listed in both syntax denotes that any three organizations endorsement 
is adequate to pass the VSCC validation. 
\par
\textbf{Observation 6:} \textit{A combination of a number of sub-policies and a number of crypto signatures
verification impacted the performance} as shown in Figure \ref{fig:endorsement}. 
The 1$^{st}$ and 4$^{th}$ \texttt{AND/OR} policies, having no sub-policies,
performed the same as \texttt{NOutOf} policies due to a few signatures
verification.
With sub-policies, 
both the number of sub-policies (i.e., search space) and the number of signatures dictated the performance. For e.g., 
the throughput achieved with 2$^{nd}$ \& 3$^{rd}$ \texttt{AND/OR} policies was 7\% \& 20\% lesser than
other policies, respectively. 
\par
\begin{figure}
    \begin{center}
     \includegraphics[scale=0.38]{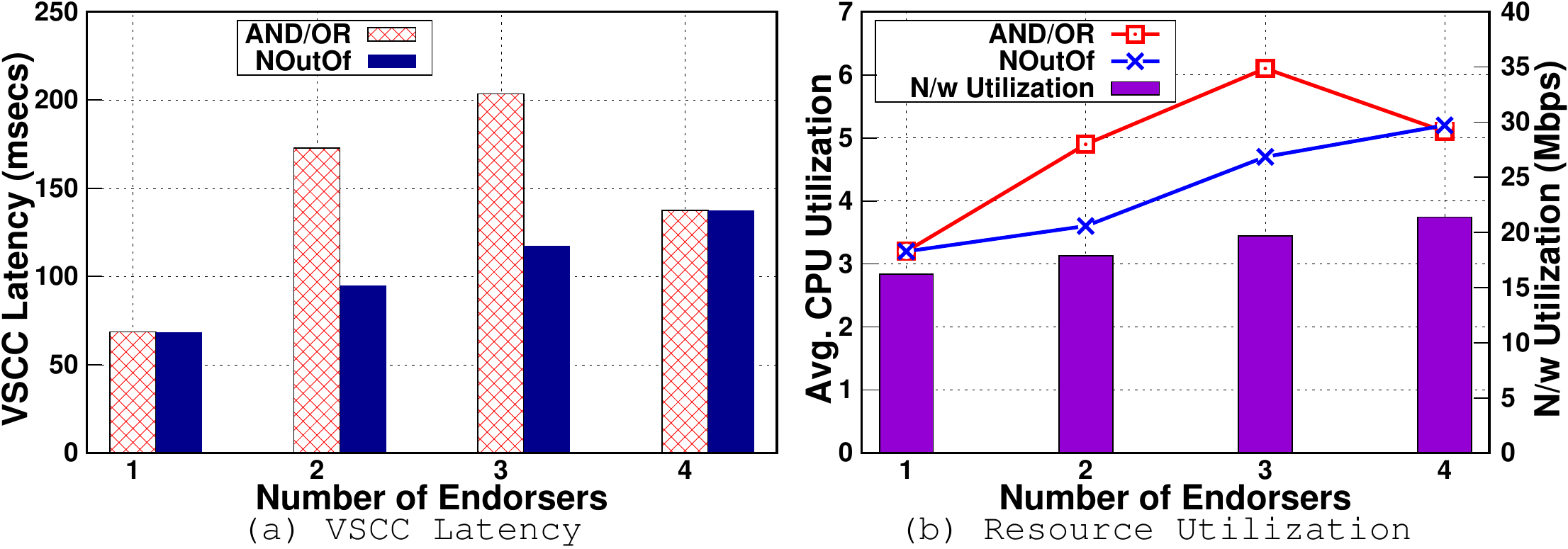}
     \vspace{-.4cm}
     \caption{VSCC latency, 
	    and resource utilization for various endorsement policies (arrival 
	    rate = 125).}
   \label{fig:endorsement-time-resources}
 \end{center}
  \vspace{-.7cm}
\end{figure}

Figure \ref{fig:endorsement-time-resources} plots the VSCC latency and 
the resource utilization at a peer node for various policies. With an increase in the number of
signatures verification (specifically for \texttt{NOutOf}), the VSCC latency increased linearly
from 68 ms to 137 ms. When there were sub-policies (as with 2$^{nd}$ and 3$^{rd}$ \texttt{AND/OR} policies), 
the VSCC latency increased significantly (i.e., 172 ms and 203 ms, respectively). A similar trend 
for resource utilization as show in Figure \ref{fig:endorsement-time-resources}(b). 
Note that the block bytes increased with increase in the number of endorsements due to the number of x.509 
certificates encoded in each transaction.

\par
There are three major CPU intensive operations during the policy validation phase (excluding the check for
satisfiability) which are listed below. 
\begin{enumerate}[itemsep=0pt, leftmargin=20pt]
	\item Deserialization of identity (i.e., x.509 certificate).
	\item Validation of identity with organization MSP \cite{msp}.
	\item Verification of signature on the transaction data.
\end{enumerate}
Hence, with an increase in the sub-policies (i.e., search space), the number of identities \& signatures 
to be validated, both CPU utilization and VSCC latency increased. It is interesting to note that the MSP 
identifier is not sent along with x.509 certificate. As a result, the policy evaluator has to validate each 
x.509 certificate with multiple organization MSPs to identify the correct one. For a 5 minutes run at an
arrival rate of 150 tps, we observed 220K such validation out of which 96K validation failed that resulted in
wastage of CPU and time.
\par
\textbf{Guideline 3:} To achieve a high performance, define policies with a fewer number of sub-policies
and signatures.
\par
\textbf{Action Item 2:} As the cryptography operations are CPU intensive, we can
avoid certain routine operations by maintaining a cache of deserialized
identity and their MSP information as shown in \S \ref{msp-cache}.  This does not introduce a security risk as
identities are long-lived and separate Certificate Revocation Lists (CRLs) are
maintained.

\subsection{Impact of Channels and Resource Allocation}
\begin{table}
 \caption{Configuration to identify the impact of channels.}

	\vspace{-.3cm}
  \label{table:config-channel}
  \begin{tabular}{  p{3.2cm} | p{4.4cm} }

        \hline
	  \textbf{Parameters} & \textbf{Values} \\ \hline
	  Number of Channels & 1, 2, 4, 8, 16 \\ \hdashline
	  Tx. Arrival Rate & 125 to 2500 tps with a step of 25 \\ \hline
  \end{tabular}
	\vspace{-0.4cm}
\end{table}

\begin{figure}[t]
    \begin{center}
     \includegraphics[scale=0.4]{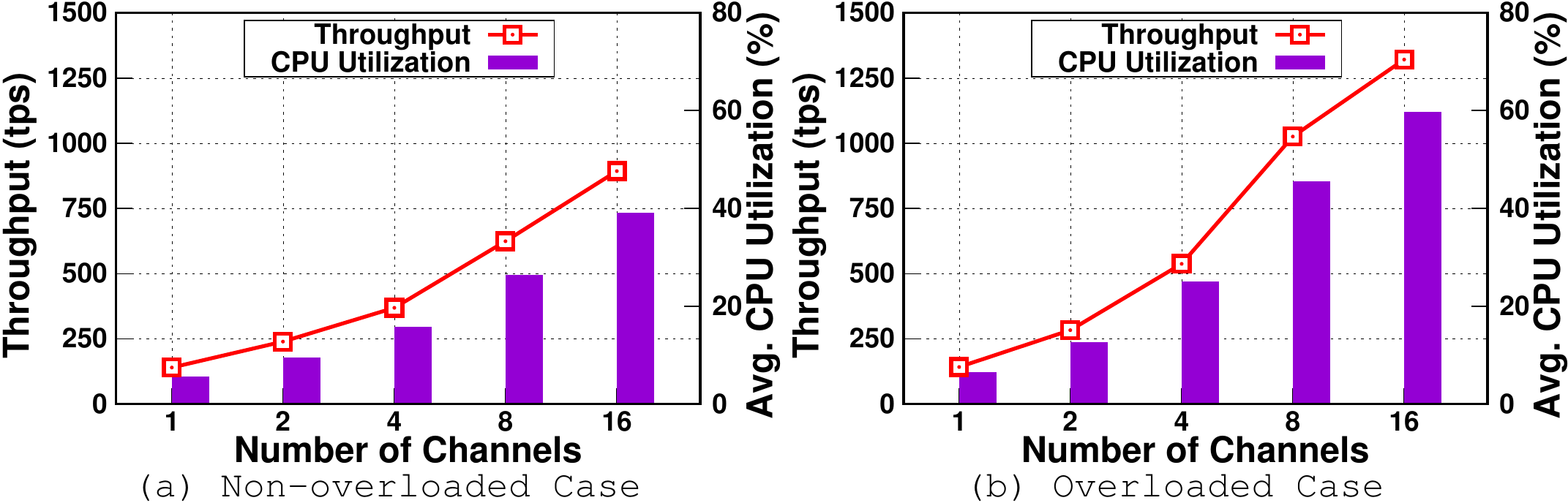}
     \vspace{-.3cm}
     \caption{Impact of the number of channels on performance.}
   \label{fig:channels}
 \end{center}
  \vspace{-.6cm}
\end{figure}
\begin{figure*}[ht]
    \begin{center}
     \includegraphics[scale=0.4]{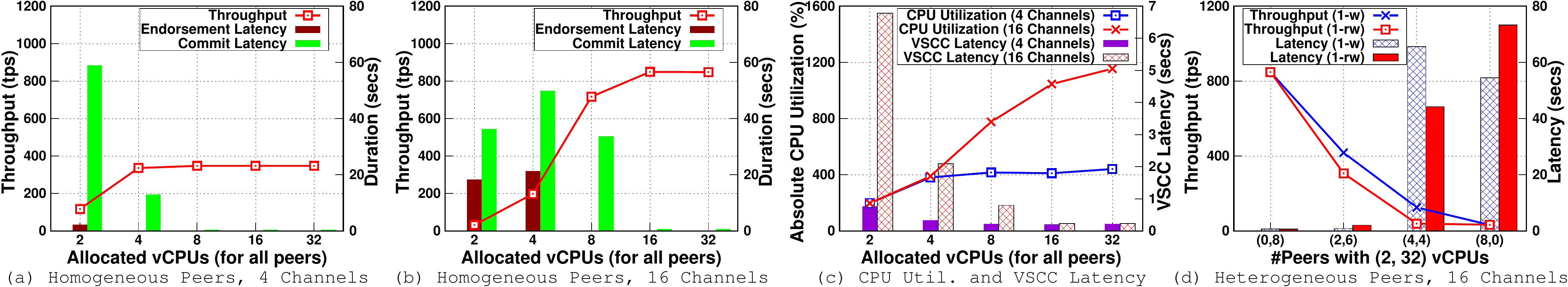}
     \vspace{-.405cm}
	    \caption{Impact of the number of vCPU on throughput \& various latencies with 4 channels (arrival rate = 350 tps) and 16 channels (arrival rate = 850 tps).}

   \label{fig:channel-resource}
 \end{center}
  \vspace{-.6cm}
\end{figure*}
We categorize the arrival rate for different channel count into two categories;
non-overloaded when the latency range was [0.4-1s] and overloaded when the latency
range was [30-40s]. 
Figure \ref{fig:channels} plots the average throughput and CPU utilization for
these two categories. Table \ref{table:config-channel} presents a various
number of channels and transaction arrival rate used for this study. 
For other
parameters,
refer to Table \ref{table:config-default}. All peers joined all the channels as 
described in \S \ref{methodology}. 
\par
\textbf{Observation 7:} \textit{With the increase in the number of channels, the throughput increased and
latency decreased. The resource utilization such as CPU also increased} as shown in Figure 
\ref{fig:channels}. For e.g., with the increase in the number of channels from 1 to 16, the throughput
increased from 140 tps to 832 tps (i.e., by 6$\times$ in non-overloaded case) and to 1320 tps (i.e., 9.5$\times$
in overloaded case). This is because each channel is independent of 
others and maintains its own chain of blocks. Hence, the validation phase and the final ledger update of multiple blocks (one per channel)
executed in parallel which increased CPU utilization that resulted in higher throughput.
\par
Figure \ref{fig:channel-resource}(a) and (b) plot the throughput, endorsement
\& commit latency for 4 and 16 channels, respectively, over a different number of allocated
vCPUs but homogeneous peers. Figure \ref{fig:channel-resource}(c) plots the absolute (instead of
average) CPU utilization across all vCPUs. Table \ref{table:config-channel-resource} presents 
a various number of vCPUs allocated, the number of channels and 
transaction arrival rate used. For other parameters, 
refer to Table \ref{table:config-default}.
\begin{figure}
    \begin{center}
     \includegraphics[scale=0.4]{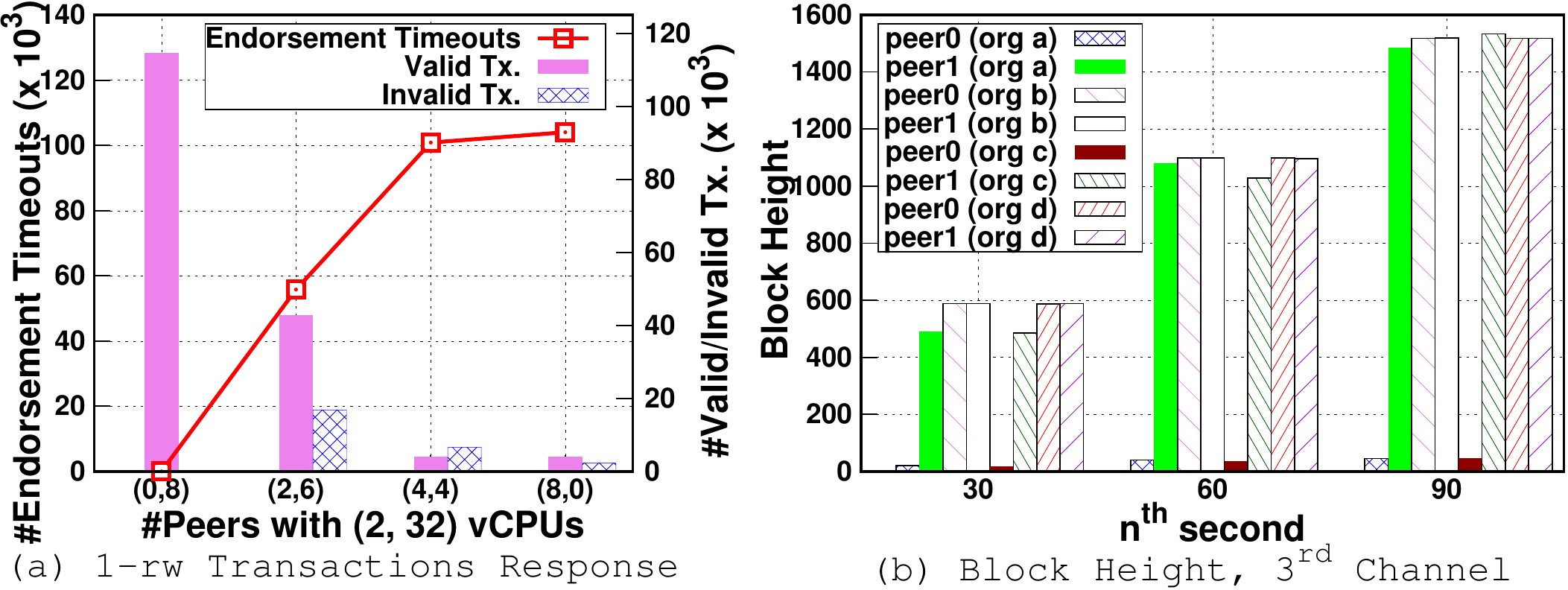}
     \vspace{-.4cm}
     \caption{Impact of heterogeneous setup.}
   \label{fig:channels-resource-hetero}
 \end{center}
  \vspace{-.6cm}
\end{figure}
\begin{table}
 \caption{Configuration to identify the impact of resource allocation.}
	\vspace{-.3cm}
  \label{table:config-channel-resource}
  \begin{tabular}{  p{4cm} | p{4cm} }

        \hline
	  \textbf{Parameters} & \textbf{Values} \\ \hline
	  Number of Channels & 4, 16 \\ \hdashline
	  Resources (same for all peers) & (2, 4, 8, 16, 32) vCPUs, 3 Gbps \\ \hdashline
	  Tx. Arrival Rate & 350 tps for 4 channels \newline 850 tps for 16 channels \\ \hline
  \end{tabular}
	\vspace{-.4cm}
\end{table}

\begin{table}[t]
 \caption{Configuration to identify the impact of heterogeneous setup.}
	\vspace{-.3cm}
  \label{table:config-channel-resource-hetero}
  \begin{tabular}{  p{2.99cm} | p{4.95cm} }

        \hline
	  \textbf{Parameters} & \textbf{Values} \\ \hline
	  Number of Channels & 16 \\ \hdashline
	  Transaction Complexity & 1 KV write (1-w) \& 1 KV read/write (1-rw) \\ \hdashline
	  \#Peers with (2, 32) vCPUs & (0, 8), (2, 4), (4, 4), (8, 0) peers \\ \hdashline
	  Tx. Arrival Rate & 850 tps \\ \hline
  \end{tabular}
	\vspace{-.4cm}
\end{table}

\par
\textbf{Observation 8:} \textit{At moderate loads, when the number of vCPUs allocated were lesser than the channel count, performance degraded}. For e.g., When the number of
vCPUs allocated were lesser than 16 for 16 channels, the throughput reduced significantly from 848 tps to 32 tps (by 26$\times$)
-- refer to Figure \ref{fig:channel-resource}(b). 
Further, both the average endorsement and commit latency exploded (from 37 ms to 21 s, 
and 640 ms to 49 s, respectively) due to a lot of contention on the CPU. 
Further, a significant number of requests got a timeout during the endorsement 
phase. Once a timeout occurs, we marked that transaction as failed. With an allocation of 
2 vCPUs, a higher number of endorsement requests got a timeout as
compared to 4 vCPUs. Thus, the endorsement and commit latency (of successful transactions) observed with 2 vCPUs 
were lesser than 4 vCPUs as shown in Figure \ref{fig:channel-resource}(b). 
\par
Due to the CPU contention, the VSCC latency also increased that affected the 
commit latency as shown in Figure \ref{fig:channel-resource}(c). 
Increasing the vCPUs allocation increased CPU utilization and consequently performance up-to a peak where 
vCPUs matched the number of channels. Beyond this, additionally allocated vCPUs were idle due to the single-threaded sequential VSCC 
validation -- refer to Figure \ref{fig:channel-resource}(c).
\par
Figure \ref{fig:channel-resource}(d) plot the throughput and
latency for 16 channels in a heterogeneous setup. Table \ref{table:config-channel-resource-hetero} presents 
a various number of vCPUs allocated for different peers to enable heterogeneity, the number of channels, transaction complexity and 
transaction arrival rate used. For other parameters, 
refer to Table \ref{table:config-default}.
\par
\textbf{Observation 9:} \textit{At moderate loads, even when the number of vCPUs allocated for 2 peers out of 8 
were lesser than the channel count, performance degraded}. For e.g., when only 2 vCPUs were allocated for 2 peers 
(others with 32 vCPUs), the throughput reduced from 848 tps to 417 tps (by 2$\times$) for write-only transactions 
and to 307 tps (by 2.7$\times$) for read-write transactions. The reasons for the reduction are twofold, 
endorsement requests timeout from less powerful peers and MVCC conflicts specifically for read/write transactions. 
Figure \ref{fig:channels-resource-hetero}(a) plots the endorsement requests timeout, valid transactions and
invalid transactions due to MVCC conflicts for read-write transactions. With an increase in the number of less 
powerful peers, the endorsement requests timeout increased. Further, a higher proportion of total submitted transactions 
became invalid due to MVCC conflicts. The is because of the lower block commit rate at
less powerful peers as compared to powerful peers. Due to the different block height at peers (refer to Figure 
\ref{fig:channels-resource-hetero}(b)), there was a mismatch of key's version in the read-set collected. As a result, 
MVCC conflicts occurred during the state validation which invalidated transactions. 
\par
Though we have not studied the impact of network resources, we believe that the impact would be similar to that of 
CPU. This is because, with a low network bandwidth, the delay in both the block and transaction delivery would increase.
Even with the heterogeneous network resources for peers, we expect the impact to be similar to the one we
observed with CPU.
\par
\textbf{Guideline 4:} To achieve higher throughput and lower latency, it is better to allocate at least one vCPU 
per channel. For optimal vCPU allocation, we need to determine the expected load at each channel and allocate
adequate vCPUs accordingly.
\par
\textbf{Guideline 5:} To achieve higher throughput and lower latency, it is better to avoid heterogeneous peers as the
performance would be dictated by less powerful peers.
\par
\textbf{Action Item 3:} Processing transactions within a channel and across channels can be improved to better utilize 
additional CPU power as shown in \S \ref{pvscc}.

\subsection{Impact of Ledger Database}
\label{impact-of-db}
\begin{figure}
    \begin{center}
     \includegraphics[scale=0.4]{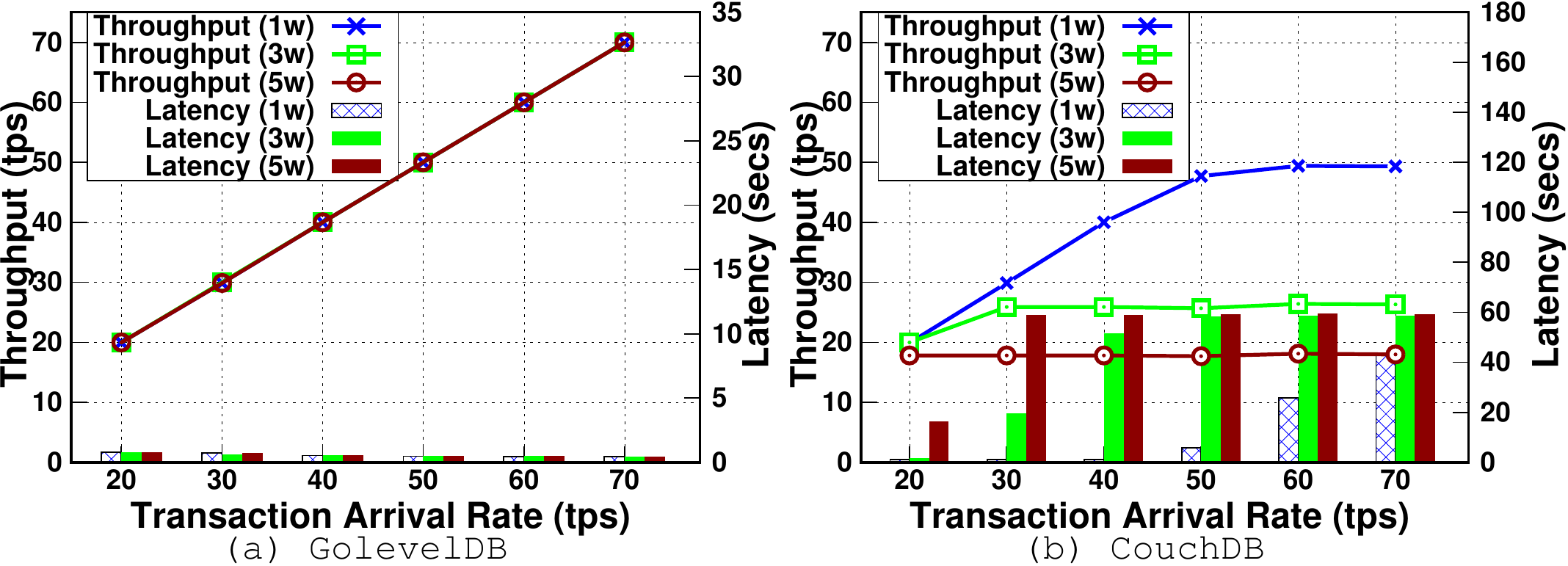}
     \vspace{-.45cm}
	    \caption{Impact of state database (one (1w), three (3w), five (5w) KV writes).}
   \label{fig:dbchoice-macro-w}
 \end{center}
  \vspace{-.5cm}
\end{figure}

\begin{figure*}
    \begin{center}
     \includegraphics[scale=0.415]{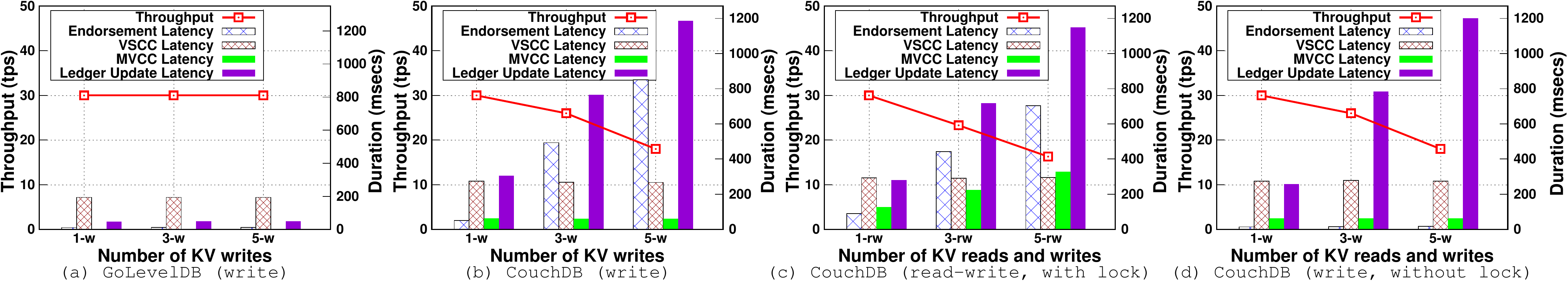}
     \vspace{-.8cm}
	    \caption{Impact of database on throughput, endorsement, VSCC, 
	    MVCC, \& ledger update latency for various transaction complexities (arrival rate = 30 tps)}
   \label{fig:dbchoice-micro-rw}
 \end{center}
  \vspace{-.5cm}
\end{figure*}
Figure \ref{fig:dbchoice-macro-w} plots the average throughput and latency over multiple transaction arrival rates for both GoLevelDB
and CouchDB with different transaction complexities. Table \ref{table:config-db} presents various databases, transaction complexity and 
arrival rate used. For other parameters,
refer to Table \ref{table:config-default}.
\par
\textbf{Observation 10:} \textit{The Fabric transaction throughput with GoLevelDB as state database was 3$\times$ greater than CouchDB}. 
The maximum throughput achieved with GoLevelDB on
a single channel was 140 tps (refer to Figure \ref{fig:block-size}) while with CouchDB, it was only 50 tps 
(refer to Figure \ref{fig:dbchoice-macro-w}(b)). 
Further, with an increase in the transaction complexity, i.e., for
multiple writes, the throughput with CouchDB dropped from 50 tps to 18 tps while no
such impact with GoLevelDB (refer to Figure \ref{fig:dbchoice-macro-w}(a)). 
The reason for significant performance differences between CouchDB and GoLevelDB 
is that the latter is an embedded database to peer process while former is accessed using REST APIs over a secure HTTP. 
As a result, the endorsement latency, VSCC latency, MVCC latency and the ledger update latency 
was higher with CouchDB as compared to the GoLevelDB as shown in Figure \ref{fig:dbchoice-micro-rw}(a) \& (b). 
\begin{table}
 \caption{Configuration to identify the impact of state database.}
	\vspace{-.3cm}
  \label{table:config-db}
  \begin{tabular}{  p{3.2cm} | p{4.4cm} }

        \hline
	  \textbf{Parameters} & \textbf{Values} \\ \hline
	  Database & GoLevelDB, CouchDB \\ \hdashline
	  Transaction Complexity & 
	  	\vspace{-.2cm}
	  	\begin{itemize}[nolistsep, leftmargin=10pt] 
			\item (1, 3, 5) KV writes
			\item (1, 3, 5) KV read/writes 
		\end{itemize}
		\\ \hdashline 
	  Tx. Arrival Rate & 20, 30, 40, 50, 60 (tps) \\ \hline
  \end{tabular}
	\vspace{-.3cm}
\end{table}

With CouchDB, the VSCC latency increased compared to the GoLevelDB as the peer accessed state database using a REST API call 
for every transaction to retrieve the endorsement policy of the chaincode on which the transaction was simulated. Similarly, the MVCC
latency also increased with CouchDB. 
\par
\textbf{Observation 11:} \textit{With CouchDB, the endorsement latency and ledger update latency increased with an increase in the number of 
writes per transaction}, i.e., from 40 ms and 240 ms with one write to 800 ms and 1200 ms with three writes, respectively, as shown in Figure \ref{fig:dbchoice-micro-rw}(b)
even though write-only transactions do not access the database during the endorsement phase. This is because the endorsement phase 
acquired a shared read lock on the whole database to provide a consistent view of data (i.e., \textit{repeatable read isolation level \cite{isolation}}) 
to the chaincode. 
Similarly, the final ledger update phase acquired an exclusive write lock on the whole database. Hence, both the endorsement phase 
and final ledger update contended for this resource. Especially, the final ledger update with CouchDB was costlier as it had to perform the following three
tasks for each key-value write in a transaction's write-set.
\begin{enumerate}[itemsep=0pt, leftmargin=20pt]
	\item Retrieve the key's previous revision number (used for concurrency control within CouchDB) by issuing a GET request, 
		if it exists in the database.
	\item Construct a document for the value (could be a JSON document or binary attachment).
	\item Update the database by submitting a PUT request.
\end{enumerate}
As a result, with the increase in the number of writes per transaction, the ledger update latency increased (refer to 
Figure \ref{fig:dbchoice-micro-rw}(b)). 
Due to the above three time consuming serial operations, we surmise that the committer held the lock on the database for a longer duration
which increased endorsement latency.
To validate our hypothesis, we performed experiments by disabling the lock acquisition on the whole database during the endorsement phase 
and final ledger update. 
The side effect 
of such action was only providing \textit{non-repeatable read isolation level} at that chaincode. As our transaction was only writing keys, 
such side effect did not affect the database consistency. Figure \ref{fig:dbchoice-micro-rw}(d) shows the improvement in endorsement phase. 
The average endorsement latency reduced from 800 ms to 40 ms, 
validating our hypothesis.
\par
\textbf{Observation 12:} \textit{Only with an increase in the number of reads per transaction, the MVCC latency increased} as shown in
Figure \ref{fig:dbchoice-micro-rw}(c). This is because with an increase in the number of items in the read set, the number of 
GET REST API calls to CouchDB increased during MVCC validation phase. With an increase in the number of writes, MVCC latency did
not increase as shown in Figure \ref{fig:dbchoice-micro-rw}(b) because it only checks whether any read keys has been modified. 
\par
\textbf{Guideline 6:} GoLevelDB is a better performant option for state
database. CouchDB is a better choice if rich-query support for read-only
transactions is important. When using CouchDB, design the
application and transaction to read/write a fewer number of keys 
to accomplish a task.
\par
\textbf{Action Item 4:} CouchDB supports bulk read/write operations \cite{bulk} without additional transactional semantics. Using
the bulk operations will reduce the lock holding duration and improve the performance as demonstrated in \S \ref{bulk}. 
\par
\textbf{Action Item 5:} The usage of database such as GoLevelDB and CouchDB,
without the snapshot isolation level, results in whole database lock during the
endorsement and the ledger update phase. Hence, our future work \cite{pgsql} is 
to look at ways to remove the lock and/or use
a database such as PostgreSQL \cite{postgres} that supports snapshot isolation.

\subsection{Scalability and Fault Tolerant} 
In Fabric, scalability can be measured in terms of the number of channels, number of organizations joining a channel and the number of peers per organization.
From a resource consumption perspective, the endorsement policy complexity controls the scalability of network. Even with 
a large number of organizations or peers, 
if the endorsement policy requires only a few organizations signature, then the performance would
be the unaffected. This is because, the transaction needs to be simulated at a fewer node in the network
to collect endorsement. Scalability could also be defined in terms of number of geographically distributed nodes
and latency in block dissemination among them. Number of ordering service
nodes and choice of consensus protocol used among them would also affect
scalability.  Though these are out of scope of this study, are important
aspects of network scalability.
\par
Node failures are common in a distributed system and hence, it is important to study the fault tolerant 
capability of Fabric. In our initial and early study, we observed that node failures do not affect the
performance (during non-overloaded case) as client can collect endorsement from other available nodes. 
With higher loads, node rejoining after a failure and syncing up the ledger due to missing blocks was observed 
to have large delays.
This is 
because though the block processing rate at the rejoined node was at the peak, other nodes continues to add 
new blocks at the same peak processing rate. 

\section{Optimizations Studied}
\label{optimizations}
In this section, we introduce three simple optimizations based on action items
listed in \S \ref{exp-results} -- (1) MSP cache in \S \ref{msp-cache},
(2) parallel VSCC validation of a block in \S \ref{pvscc}, and (3) bulk read/write\footnote{A Fabric community 
proposal and an implementation of bulk operation API was available but was not integrated with
MVCC and the final commit} during MVCC validation \& commit for CouchDB in \S \ref{bulk}.
For
each of these optimizations, first, we study the performance improvement individually. Then,
we study the improvement by combining all the three optimizations. 
\subsection{MSP Cache}
\label{msp-cache}
\begin{table}[ht]
\begin{minipage}[b]{0.5\linewidth}
	\centering
  \label{table:config-msp-cache}
  \begin{tabular}{  p{1.3cm} | p{1.1cm} }
        \hline
	  \textbf{Parameters} & Values \\ \hline
	  {Endorsement Policy} (\texttt{AND}/\texttt{OR})& all four from Table \ref{table:config-endorsement}\\ \hdashline
	  {Tx. Arrival Rate}& 400, 500, 600 (tps)\\ \hline
  \end{tabular}
 \caption{Configuration to identify the efficiency of MSP cache.}

\end{minipage}\hfill
\begin{minipage}[b]{0.5\linewidth}
     \includegraphics[scale=0.41]{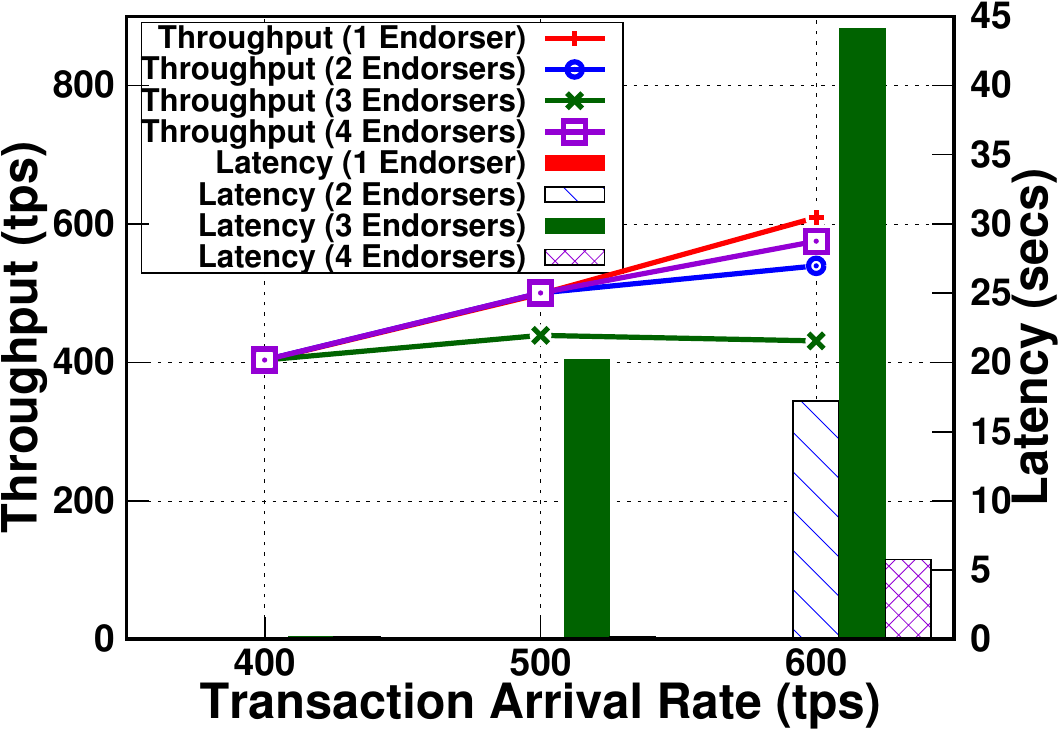}
     \vspace{-.6cm}
	    \captionof{figure}{Impact of MSP cache.}
   \label{fig:endorsement-msp}
\end{minipage}
\end{table}

\begin{figure}[t]
    \begin{center}
     \includegraphics[width=8cm,height=4.5cm]{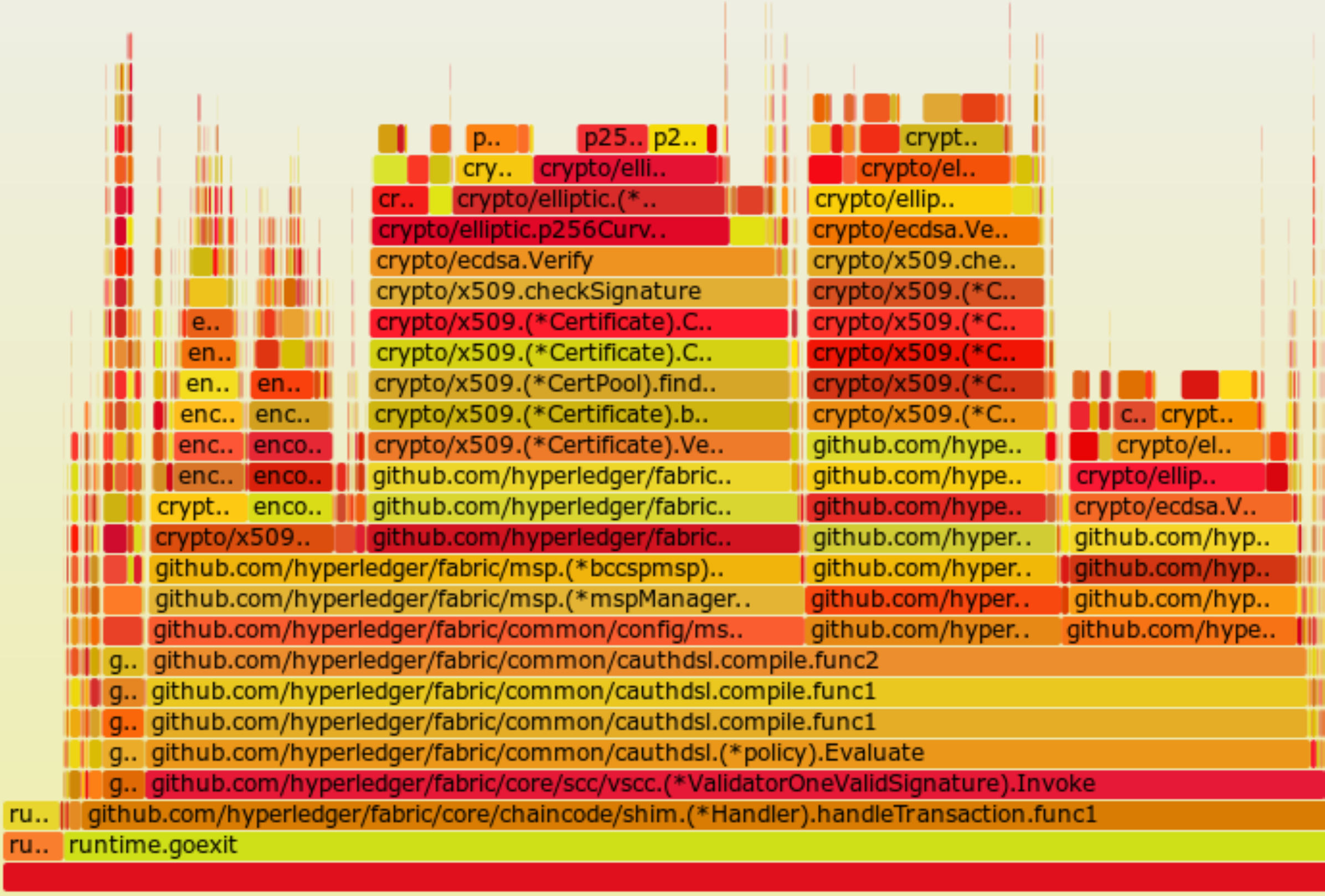}
     \vspace{-.4cm}
	    \caption{Frequency and call stack depth of crypto operations (3$^{rd}$ policy in 
	    \texttt{AND/OR}) -- without the MSP cache}
   \label{fig:endorsement-flame}
 \end{center}
  \vspace{-.3cm}
\end{figure}
\begin{figure}[t]
  \vspace{-.2cm}
    \begin{center}
     \includegraphics[scale=0.35]{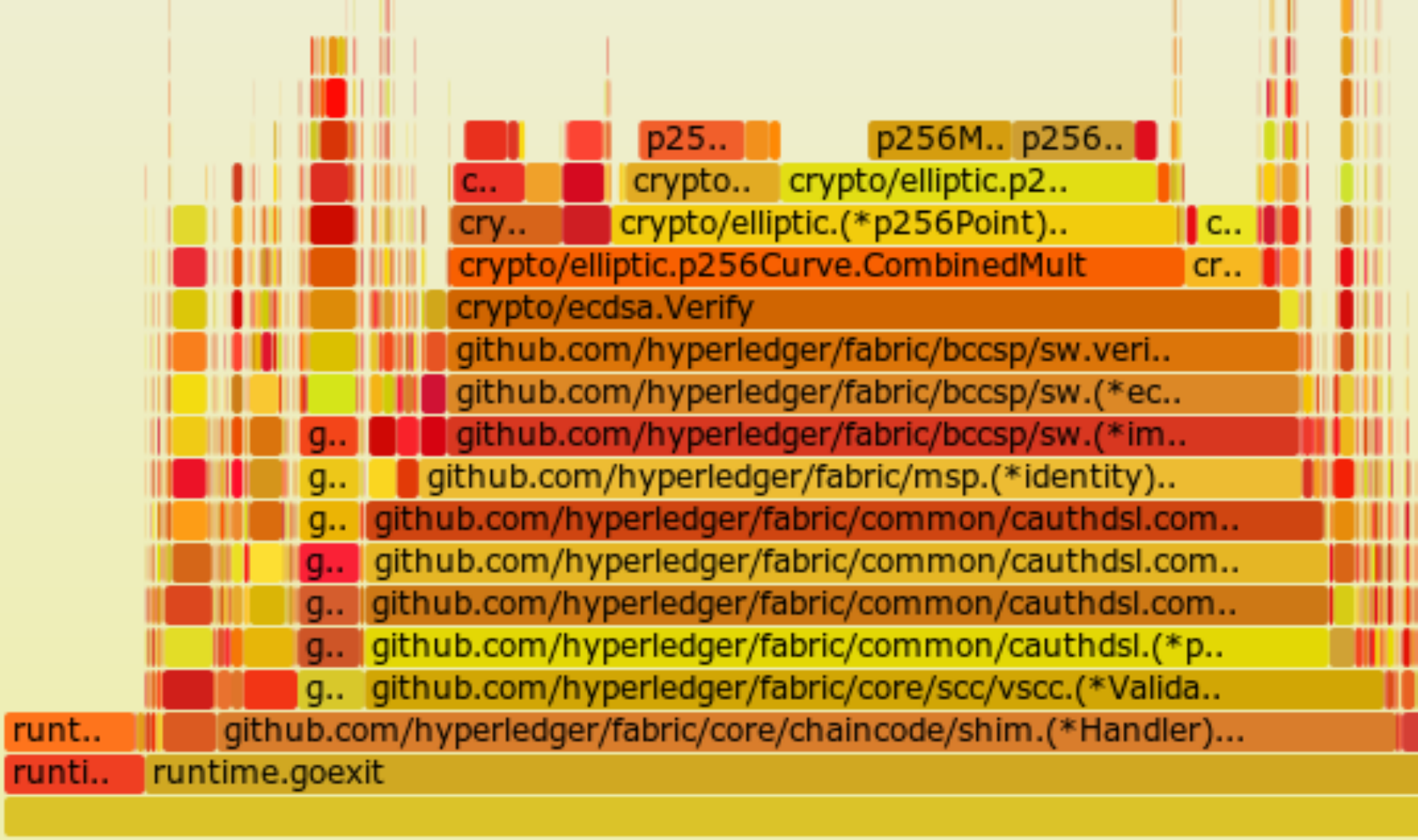}
     \vspace{-.4cm}
	    \caption{Frequency and call stack depth of crypto operations (3$^{rd}$ policy in \texttt{AND/OR}) 
	    -- with the MSP cache.}
   \label{fig:endorsement-msp-flame}
 \end{center}
  \vspace{-.5cm}
\end{figure}
As crypto operations are very CPU intensive, in this section, we studied the efficiency of using a cache at the 
following two operations in the crypto module:  
\begin{enumerate}[nolistsep, leftmargin=20pt]
	\item Deserialization of identity (i.e., x.509 certificate).
	\item Validation of identity with Organization's MSP.
\end{enumerate}
To avoid deserialization of the serialized identity every time, we cached the deserialized identity using a hash map with the 
serialized form as key. Similarly, to avoid validating an identity with multiple MSPs every time, we used
a hash map with a key as identity and value as the corresponding MSP to which the identity belongs. Further, we 
employed the ARC \cite{arc} algorithm for cache replacement. During identity revocations, we invalidated cache entries
appropriately. 
\par
Figure \ref{fig:endorsement-msp} plots the impact of MSP cache on the throughput and latency for \texttt{AND/OR} endorsement
policies over different transaction arrival rates. Table \ref{table:config-endorsement} presents various policies 
used along with different transaction arrival rates. We draw
the attention of the reader to Figure \ref{fig:endorsement}(b) for comparison against the no-cache
behavior. On average, the throughput increased by 3$\times$ due to MSP cache
as compared to a vanilla peer. For e.g., when the endorsement policy required
signature from two endorsers (defined using \texttt{AND/OR} 
syntax), the maximum throughput achieved without MSP cache was 160 tps while with cache, it increased to 540 tps. This is because 
the MSP cache reduced certain repetitive CPU intensive operations.
\par
Figure \ref{fig:endorsement-flame} and 
\ref{fig:endorsement-msp-flame} plots the flame graph showing frequency of crypto operations and call stack depth of VSCC validation phase in a 
vanilla peer and a peer with MSP cache, respectively. As it can be observed, the number of crypto operations and call stack depth 
reduced significantly with the MSP cache. 

\subsection{Parallel VSCC Validation of a Block}
\label{pvscc}
\begin{figure}
    \begin{center}
     \includegraphics[scale=0.39]{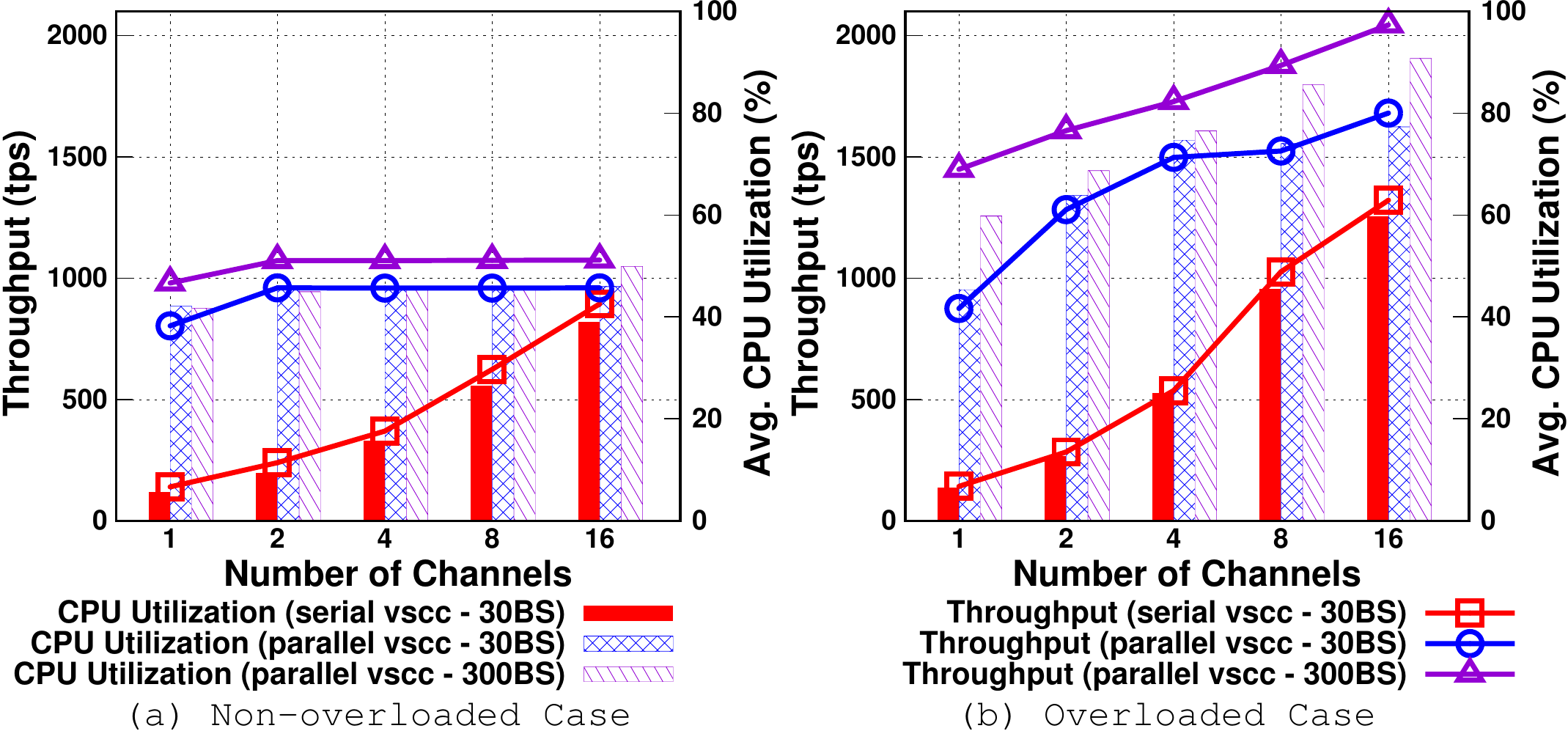}
     \vspace{-.3cm}
     \caption{Impact of parallel VSCC validation on multichannel setup.}
   \label{fig:pvscc}
 \end{center}
  \vspace{-.4cm}
\end{figure}
\begin{figure*}[h]
    \begin{center}
     \includegraphics[scale=0.405]{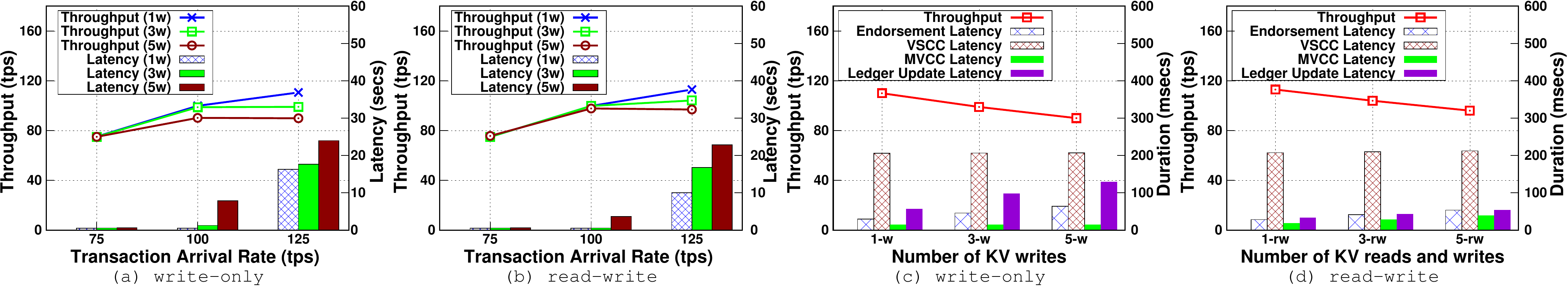}
     \vspace{-.4cm}
     \caption{Impact of bulk read during the MVCC validation and ledger update on the performance.}
   \label{fig:couch-batching}
 \end{center}
  \vspace{-.5cm}
\end{figure*}

The VSCC validation phase validates each transaction in a block serially 
against the endorsement policy. As this approach under-utilized the resources, we studied the
efficiency of parallel validation, i.e., validate multiple transactions' endorsement 
in parallel to utilize otherwise idle CPU and improve the overall performance. 
To achieve this, we created a configurable number of worker threads per channel on
peer startup. Each worker thread validates one transaction's 
endorsement signature set against its endorsement policy. 
\par
Figure \ref{fig:pvscc} plots the impact of parallel VSCC on the performance
and resource utilization. We categorize the arrival rate for different channel count 
into two categories; non-overloaded case when the latency falls in [0.1-1s] and overloaded 
when the latency falls in [30-40s]. For each channel, we allocated worker
threads equal to the block size. The throughput and resource utilization during the non-overloaded case
for one channel exploded from 130 tps to 800 tps (improved by 6.3$\times$) for the
block size of 30 and to 980 tps (7.5$\times$) for the block size of 300. 
This is due to parallel validation and hence the reduction in the VSCC latency (from 300 ms to 30 ms, i.e., by 10$\times$ 
reduction for the block size of 30). The throughput saturated at 950 tps (for the block size of 30) \& 
1075 tps (for the block size of 300) -- refer to Figure \ref{fig:pvscc}(a). 
\par
Similarly, the throughput and resource
utilization during the overloaded case increased to 1.5$\times$ for 16 channels to as much as 10$\times$ for 1 channel -- refer
to Figure \ref{fig:pvscc}(b). 
This shows that parallel VSCC validation of a 
block significantly increases the performance of a single channel.
With an increase in the number of channels, the percentage of improvement decreased. 
This is because multiple channels by default result in the parallel validation of blocks (instead of transactions) and hence 
a few number of free vCPUs were available for parallel VSCC.

\subsection{Bulk Read/Write During MVCC Validation \& Commit}
\label{bulk}
\begin{figure*}
    \begin{center}
     \includegraphics[scale=0.36]{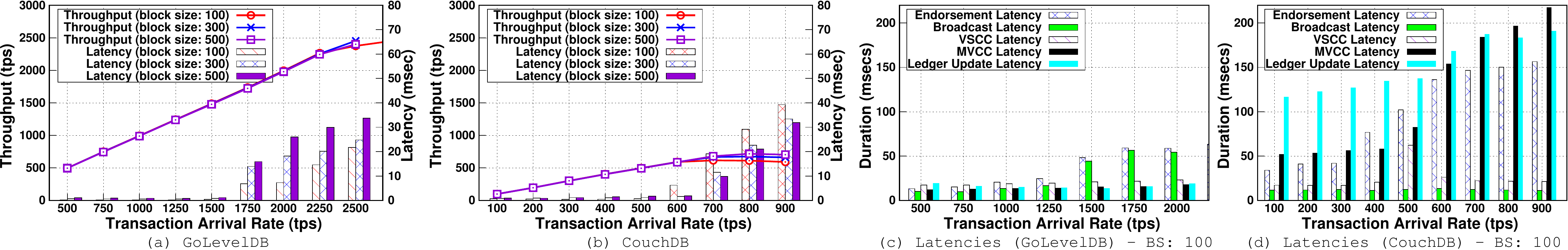}
     \vspace{-.3cm}
     \caption{Impact of all the three optimizations on the performance with different block sizes.}
	\label{fig:all-opt}
 \end{center}
  \vspace{-.7cm}
\end{figure*}

During the MVCC validation, with CouchDB as the state database, for each transaction in a block, for each key in the
read set of the transaction, a GET REST API call to the database over a secure HTTPS retrieved the last committed version number. 
During the commit phase, for each valid transaction (recorded after MVCC validation) in a block, for each key in the write 
set of the transaction, a GET REST API 
call retrieved the revision numbers \cite{rev}. Finally, for each entry in the write set,
a PUT REST API call committed the document. Due to these multiple REST API calls, 
performance degraded significantly as demonstrated in \S \ref{impact-of-db}. 
\par
To cut down the number of REST API calls, CouchDB suggests using bulk
operations. 
Hence, we used 
the existing
\texttt{BatchRetrieval} API in Fabric to batch load multiple keys' version and revision number into
the cache over a single GET REST API call per block. 
To enhance the ledger update process, we used \texttt{BatchUpdate} API in Fabric to commit a batch of documents using 
a single PUT REST API call per block.
Further, we introduced a cache in VSCC
to reduce the calls to CouchDB to obtain the endorsement policy of
the chaincode for each transaction.  In this section, we show the efficiency of these enhancements
on the overall performance.  
\par
Figure \ref{fig:couch-batching} plots the throughput and latency when running a CouchDB as the state database
with the bulk read/write optimization. For comparison against the non-bulk read/write, refer to Figure \ref{fig:dbchoice-macro-w}(b) 
and Figure \ref{fig:dbchoice-micro-rw}. 
The performance increased significantly from 50 tps
to 115 tps (i.e., by 2.3$\times$) for transactions with a single write. For multiple writes (3-w \& 5-w),
the throughput increased from 26 tps to 100 tps (i.e., 3.8$\times$ for 3-w), and 18 tps to 90 tps (i.e., 5$\times$ for 5-w). 
We noticed similar improvements for read-write transactions. 
\par
Due to the bulk read/write 
optimization, the MVCC latency, ledger update latency and endorsement latency
decreased as shown in Figure \ref{fig:couch-batching}(c) and (d) as compared to Figure \ref{fig:dbchoice-micro-rw}. 
The reduction in endorsement latency (by at least 3$\times$) 
was because of the reduction in lock holding duration by the commit phase (by at least 8$\times$). 
The MVCC latency for read-write transactions reduced (by at least 6$\times$) due to a bulk 
reading of all keys in the read set of all transactions in a block. Note that the MVCC latency increased with the increase in the number of
keys read in a bulk read. The ledger update latency of a block encompassing a higher number of write-only transactions was higher. 
This is because, in read-write transactions, the MVCC validation phase itself loaded the required 
revision numbers into the cache (as the transaction read those keys before modification) which was not the case 
with write-only transactions. 

\subsection{Combinations of Optimizations}
\begin{table}
	\vspace{-.2cm}
 \caption{Configuration to identify the impact of all three optimizations combined.}
	\vspace{-.3cm}
  \label{table:config-all-opts}
  \begin{tabular}{  p{4cm} | p{3.6cm} }

        \hline
	  \textbf{Parameters} & \textbf{Values} \\ \hline
	  Number of Channels & 1, 8, 16 \\ \hdashline
	  Transaction Complexity & 1 KV write (1-w) of 20 bytes \\ \hdashline
	  StateDB Database & GoLevelDB, CouchDB \\ \hdashline
	  Peer Resources & 32 vCPUs, 3 Gbps link \\ \hdashline
	  Endorsement Policy  &  $1^{st}$ and $3^{rd}$ \texttt{AND/OR} policies \\ \hdashline
	  Block Size & 100, 300, 500 (\#tx) \\ \hdashline
	  \#VSCC Workers per Channel & Equal to the block size\\ \hline
  \end{tabular}
	\vspace{-.4cm}
\end{table}

Figure \ref{fig:all-opt}
plots the performance improvement achieved with all three optimizations combined.
Table \ref{table:config-all-opts} presents the number of VSCC worker threads per
channel, block sizes, and other relavant parameters used for this study.
\par
With GoLevelDB as the state database, the single channel throughput increased to 2250 tps from 140 tps 
(i.e., 16$\times$ improvement) due to all three optimizations -- refer to Figure \ref{fig:all-opt}(a)
and Figure \ref{fig:block-size}. Similarly, with CouchDB as 
the state database, the singe channel throughput increased to 700 tps from 50 tps (i.e., 14$\times$ 
improvement) -- refer to Figure \ref{fig:all-opt}(b) and Figure \ref{fig:dbchoice-macro-w}. 
With an increase in the block size, when CouchDB was the state database, we observed a
lower total latency due to the reduction in the number of bulk REST API call to CouchDB (i.e, 
for 500 transactions, only 2 bulk REST API calls, one read call during MVCC phase and one write
during commit phase, were issued when the block size was 500 as compared to 10 bulk REST API calls for a 
block size of 100). As a result, our guideline 1 \& 2 are not applicable for CouchDB with bulk read/write optimizations. 
\par
\begin{figure}
    \begin{center}
     \includegraphics[scale=0.33]{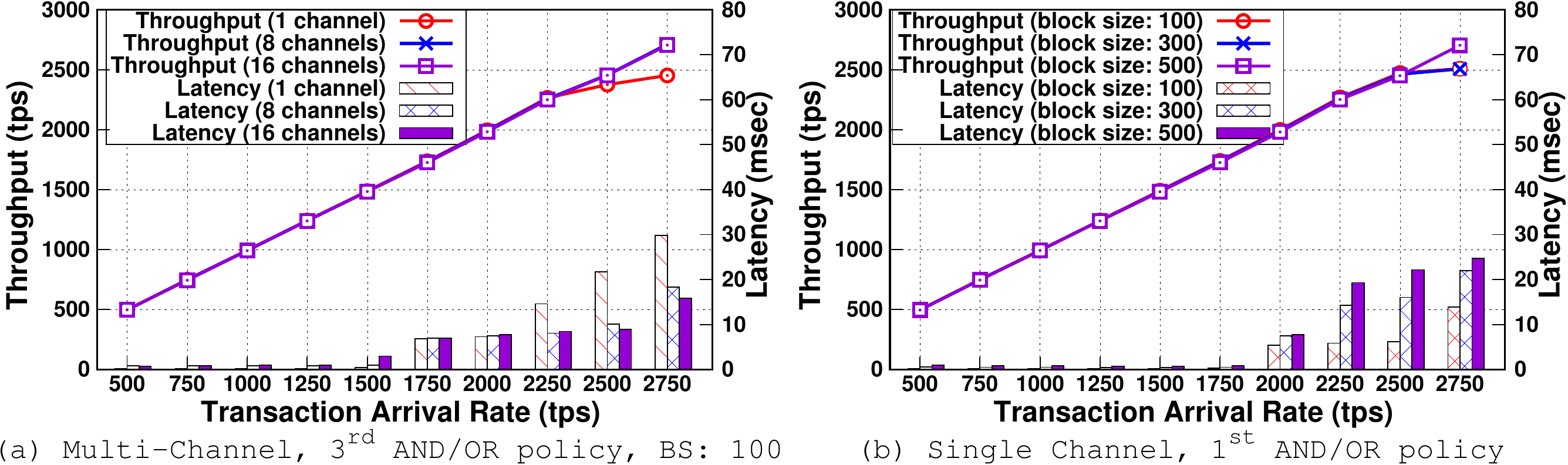}
     \vspace{-.5cm}
     \caption{Impact of all the three optimizations on the performance with a simple endorsement policy and
	    different number of channels.}
	\label{fig:all-opt-channel}
 \end{center}
  \vspace{-.7cm}
\end{figure}

Further, for 8 and 16 channels, the throughput increased to 2700 tps from 1025 tps and 1321 tps, respectively
as shown in Figure \ref{fig:all-opt-channel}(a). 
With a simpler endorsement policy, i.e., $1^{st}$ \texttt{AND/OR} policy, 
the single channel throughput also increased to 2700 tps (with GoLevelDB) as shown in Figure \ref{fig:all-opt-channel}(b).
 \par
Even with a throughput of 2700 tps, the average CPU utilization of a peer was only 60\% and the network utilization of a peer
was 1680 Mbps (send) and 240 Mbps (receive). This is because  
the sum of MVCC latency and ledger update latency (less CPU intensive tasks) was almost same or higher than the VSCC latency
(as shown in Figure \ref{fig:all-opt}(c) and (d)). Due to these sequential phases, vCPUs were underutilized.
One potential optimization would be to pipeline the VSCC and MVCC validation phase.

\section{Related Work}
\label{related-work}
There has been considerable interest in the scalability and performance
characteristics of public blockchain networks and specifically the limiting
factor of the consensus protocol and its security implications
\cite{pow-bft, gervais}.
\par
Also for public blockchains, \cite{croman2016scaling} have looked at quantifying
throughput, latency, bootstrap time and cost per transaction for the
Bitcoin network based on publicly available data.
\par
BlockBench \cite{blockbench} was one of the first to look at permissioned
blockchain. They present a framework for comparing performance of different
blockchain platforms, namely, Ethereum, Parity and Hyperledger Fabric using a
set of micro and macro benchmarks. Similar to \cite{croman2016scaling} they
generalize consensus, data, execution and application as 4 layers of blockchain
and use the benchmarks to exercise them. They measure the overall performance
in terms of throughput, latency and scalability of the platforms and draw
conclusions across the 3 platforms. However, they studied the performance of Fabric v0.6, 
with v1.0 version bringing in a complete re-design, their
observations do not hold relevance and needs re-study.
\par
\cite{fabcoin} presents the design and the new architecture of
Fabric, delving in-depth into its design considerations and modularity. It
presents the performance of a single Bitcoin like crypto currency application on
Fabric, called Fabcoin, which uses a customized VSCC to validate 
the Fabcoin specific transactions and avoid complex endorsements and channels. 
Further, they used CLI command to emulate clients, which is not realistic, 
instead of using a SDK \cite{sdk-go, sdk-java, sdk-node}. Our work differs from theirs in that, we do a comprehensive 
study for different workloads keeping Fabric's modularity and application 
in multiple domains in focus.
\par
A note to the reader, \cite{fabcoin} used Fabric v1.1-preview release, 
which incorporates all our optimizations and other additional functionalities
over v1.0.  However, being a minor version update much of the core
functionality remains the same and our observations hold true for v1.1 and
future versions based on the new architecture of Fabric.

\section{Conclusion \& Future Work}
\label{conclusion}
In this paper, we conducted a comprehensive empirical study to understand 
the performance of Hyperledger Fabric, a permissioned blockchain platform, by varying values assigned 
to configurable parameters such as block size, endorsement policy,
channels, resource allocation, and state database choices. As
a result of our study, we provided six 
valuable guidelines on configuring these parameters and also identified three
major performance bottlenecks. Hence, we introduced and studied three simple
optimizations such as MSP cache, parallel VSCC validation, and
bulk read/write during MVCC validation \& commit phase to improve the singe
channel performance by 16$\times$. Further, these three optimizations 
have been successfully adopted in Fabric v1.1
\par
As a part of future work, we will study the scalability and
fault tolerant capability of Fabric by using different blockchain 
topologies such as different number of organizations and different 
number of nodes per organization. Further, we plan to 
quantify the impact of various consensus algorithms and number of 
nodes in the ordering service on the 
performance of different workloads. In our study, we assumed that 
the network is not a bottleneck. However, in the real world setup, 
nodes can be geographically distributed and hence, the network might 
play a role. In addition, the arrival rates in real world production 
system would be following certain distributions. Hence, we will
study the performance of Fabric in Wide Area Network (WAN) with 
different arrival rate distributions.

\section{Acknowledgements}
We wish to acknowledge our following colleagues for their valuable assistance
to our work.  Our proposed optimizations were successfully adopted to Fabric,
thanks to fabric developers who took care of submitting patches to Fabric v1.1.
Angelo De Carlo (MSP Cache), Alessandro Sorniotti (Parallel Validation). Thanks
to David Enyeart, Chris Elder, Manish Sethi for their proposal on Bulk Read from 
CouchDB. We would like to thank Yacov Manevich for his consistent help. 

\bibliographystyle{abbrv}
{
	\balance
\bibliography{ms}
}
\end{document}